%% file: disk_ko.tex
\documentclass[preprint]{aastex}

\newcommand\cs{c_s}
\newcommand\vA{v_A}
\newcommand\kx{{k_x}}
\newcommand\ky{{k_y}}

\newcommand\Msun{{\rm\,M_\odot}}

\newcommand\pc{{\rm\,pc}}

\newcommand\alf{Alfv\'en }
\newcommand\torb{t_{\rm orb} }

\newcommand\simgt{\lower.5ex\hbox{$\; \buildrel > \over \sim \;$}}
\newcommand\simlt{\lower.5ex\hbox{$\; \buildrel < \over \sim \;$}}

\shortauthors{Kim \& Ostriker}
\shorttitle{Gravitational Instabilities in Disks}

\begin{document}

\pagebreak
\title{Amplification, Saturation, and $Q$ Thresholds for Runaway: \\
Growth of Self-Gravitating Structures in Models of \\
Magnetized Galactic Gas Disks}

\slugcomment{Accepted for publication in the Astrophysical Journal
\hspace{1.2cm}}

\author{Woong-Tae Kim and Eve C. Ostriker}
\affil{Department of Astronomy, University of Maryland \\
College Park, MD 20742-2421}

\email{kimwt@astro.umd.edu, ostriker@astro.umd.edu}

\begin{abstract}
We investigate the susceptibility of gaseous, magnetized galactic
disks to formation of self-gravitating condensations using
two-dimensional, local models.  We focus on two issues: (1)
determining the threshold condition for gravitational runaway, taking
into account nonlinear effects, and (2) distinguishing the
magneto-Jeans instability (MJI) that arises under inner-galaxy
rotation curves from the modified swing amplification (MSA) that
arises under outer-galaxy rotation curves.  For axisymmetric density
fluctuations, instability is known to require a Toomre parameter
$Q<1$.  For nonaxisymmetric fluctuations, any nonzero shear $q\equiv
-d\ln \Omega /d \ln R$ winds up wavefronts such that in
linear theory amplification saturates.  Any $Q$ threshold for
nonaxisymmetric gravitational runaway must originate from nonlinear
effects.  We use numerical magnetohydrodynamic simulations to
demonstrate the anticipated threshold phenomenon, to analyze the
nonlinear processes involved, and to evaluate the critical value $Q_c$
for stabilization. We find $Q_c\sim 1.2-1.4$ for a wide variety of
conditions, with the largest values corresponding to nonzero but
subthermal mean magnetic fields.  Our findings for $Q_c$ are similar
to those inferred from thresholds for active star formation in the
outer regions of spiral galaxies.  MJI is distinct from MSA in that
opposition to Coriolis forces by magnetic tension, rather than
cooperation of epicyclic motion with kinematic shear, enables
nonaxisymmetric density perturbations to grow.  We suggest that under
low-shear inner-disk conditions, $Q_c$ will be larger than that in
outer disks by a factor $\sim (v_A/q c_s)^{1/2}$, where $v_A$ and
$c_s$ are the respective Alfv\'en and sound speeds.
\end{abstract}

\keywords{galaxies: ISM --- galaxies: kinematics and dynamics
--- galaxies: structure --- instabilities --- ISM: kinematics and dynamics
--- ISM: magnetic fields --- MHD --- stars: formation}

\section{Introduction}

\subsection{Observational motivation and previous theory}

In the Milky Way and external spiral galaxies, most of the molecular gas 
is found in cold, giant molecular clouds (GMCs) within which most of star 
formation takes place (e.g., \cite{you91,wil00}). 
Star forming GMCs are strongly associated with spiral arms 
(e.g., \cite{sta79, sol85, kenney97}), and often tend to appear in clusters, 
forming giant molecular associations (GMAs) 
(e.g., \cite{coh85, gra87, vog88, ran90, ran93, sak99}). 
GMAs are present even in 
flocculent galaxies where spiral arms are relatively 
weak \citep{sak96, tho97a, tho97b}. Recent studies of internal structure and 
dynamics have significantly improved our understanding of GMCs and GMAs;
but their formation mechanisms remain subject to considerable uncertainty.

A suggestion of long standing is that GMCs or GMAs form through
gravitational instabilities either in galactic disks at large or within 
spiral arms (\cite{gol65, lyn66, elm79, elm87a, elm94, elm83, bal85, bal88,
gam92, gam96}, see also reviews in \cite{elm93a, elm95a, elm96} and 
references therein). \citet[hereafter GLB]{gol65} used a linear-theory 
analysis to show that nonaxisymmetric perturbations may grow strongly in a 
shearing disk, provided self-gravity is sufficient. Amplification of a 
linear disturbance is ultimately limited 
by differential rotation, which shears any wavelet into a tightly-wrapped 
trailing spiral pattern in which (stabilizing) pressure overmasters
(destabilizing) self-gravity. 
This mechanism for transient growth of perturbations in both the gaseous 
disks of GLB and their stellar counterparts \citep[hereafter JT]{jul66} 
has come to be called a ``swing amplifier'' \citep{too81}.

Operating on a large scale in a galactic disk, swing amplification
may potentially lead to the formation of self-gravitating
cloud complexes with sizes and masses set by the two-dimensional (2D)
Jeans scale \citep{elm83, elm87a}. The characteristic Jeans-unstable
scale of $\sim 1\,\rm kpc$ and mass of $\sim10^7\,\Msun$ (see \S2.2)
appear to be in good agreement with observed HI superclouds
\citep{elm83, elm87b, kna93} 
or GMAs \citep{ran93, sak96, tho97a, tho97b, sak99}.
Gravitational instabilities may also operate on a smaller scale
within a pre-existing spiral arm \citep{bal85, bal88, elm94}.
In particular, \citet{bal85} and \citet{bal88} showed that the 
density enhancement and shear profile inside a spiral wave crest 
can trigger near-axisymmetric and 
swing-like gravitational instabilities that could 
develop into cloud complexes with masses and spacings 
consistent with observations. 
The presence of spiral arms may not be necessary for cloud formation
via instabilities, but the ambient high-density, low-shear conditions 
within arms are generally more favorable for Jeans-type instabilities than
conditions in interarm regions \citep{elm94}.

Effects of magnetic fields on linear gravitational instabilities 
in 2D differentially-rotating disks have been thoroughly investigated 
by \citet{elm87a, elm94}, \citet{gam96} and \citet{fan97}.
These studies extended the works of \citet{cha54} and \citet{lyn66}
for gravitational instability in a magnetized, uniformly-rotating disk 
to incorporate the effects of shear analyzed by GLB. 
In regions of 
disks with weak shear, magnetic field tension enhances instability
by redistributing angular momentum (breaking the conserved-potential
vorticity constraint), while in strongly-shearing regions the 
magnetic field pressure combines with
thermal pressure (and the Coriolis force) in tending to suppress 
linear instabilities.

Formation of molecular clouds via large-scale gravitational instability
also, implicitly, underlies the proposal that global star formation
in disk galaxies is regulated by the value of the Toomre $Q$
parameter (for the definition of $Q$ see eq.\ [\ref{TQ}] below).
Following the pioneering work of \citet{qui72}, who suggested
a threshold model for star formation, \citet{ken89} found that 
active star formation requires azimuthally-averaged 
gas surface density corresponding to $Q\simlt Q_c=1.6$ 
for massive spirals\footnote{The enhanced study of \cite{mar01}, contemporary
with the present work, reports a mean observational threshold for active
star formation at $Q_c=1.4$.}
(also see the review of \cite{ken97} and references therein).
This azimuthally averaged $Q$ threshold model appears successfully to explain 
the radial distributions of star formation rates and
gas surface densities also in low surface brightness spirals
\citep{van93} and  HII galaxies \citep{tay94}.
\citet{ran93} studied the azimuthal variation of
surface density in the grand-design spiral M51 and found that
the ratio of actual to critical surface densities
is higher in the arms than in the interarm regions, indicating
that arm gas is more prone to the gravitational instabilities.

While models that appeal to a threshold in the Toomre $Q$ parameter 
as a star formation criterion are operationally quite successful in 
identifying observed outer radii for active star formation in spiral galaxies, 
the theoretical underpinning of the threshold concept is indirect.
In particular, the literature has lacked an explicit theoretical
demonstration and evaluation of thresholds for generic ({\it nonaxisymmetric})
gravitational runaways.
The star formation threshold concept, by utilizing an azimuthally-averaged 
surface density, borrows its physical grounds from the 
{\it axisymmetric} instability of \citet{saf60} and \citet{too64}.
Axisymmetric (ring-like) instability in a thin gas disk requires $Q<1$
(\cite{lin71}; see e.g., \cite{shu92}), which is generally 
not realized in galaxies. \citet{ken89} attributed the 
fact of the apparent critical $Q_c$ exceeding unity
to the results of two-fluid (star+gas) instabilities \citep{jog84a, jog84b},
with the stellar disk's self-gravity enhancing instability and hence
raising $Q_c$. The stabilizing effect from pervasive magnetic fields to
axisymmetric perturbations, which would tend to decrease $Q_c$, 
must however be considered as well \citep{elm92}.

In the more general, {\it nonaxisymmetric} case, perturbations experience
significant amplification as they swing from leading to trailing 
(GLB; JT; \citet{too81}) for a range of $Q$ both greater and 
smaller than unity. Because the magnitude of swing amplification is
continuous with respect to $Q$ in the linear theory, those analyses
cannot establish a decisive critical value for $Q$.
\citet{elm91, elm94} suggested that parallel secondary instabilities
in either shearing wavelets that have grown from swing amplification,
or in spiral density waves, have effective thresholds $Q_c\sim1-2$. 
But this suggestion needs to be confirmed by numerical simulations,
because the background state for onset of parallel instabilities is 
not in the linear regime.

Although the linear gravitational instability analyses are useful and
effective in predicting approximate sizes, spacings, and 
formation time scales of giant cloud complexes, as well as 
approximate star formation threshold densities and radii, the linear theory  
still has to be supplemented by nonlinear simulations. This is because the
instability in either general disk planes  or spiral arms shows 
only transient growth, eventually stabilized
by the kinematic increase of the local wavenumber due to background 
shear (e.g., \citet{elm87a}) or by
expansion of the background flow off the arms \citep{bal85}.
The transient nature of local gravitational instabilities 
allows only limited time for perturbations 
to grow. The evolution of a system subject to a transient 
instability depends on the initial perturbation amplitude;
if the initial perturbation level is too low, the system may
never reach a nonlinear phase.
This is unlike a true instability, in which disturbances grow exponentially 
until nonlinear saturation occurs, regardless of their initial amplitudes. 
The amplification factor in shear instability is a sensitive
function of wavenumbers, and mode coupling is inevitable in the nonlinear
stage of evolution. Numerical simulations with realistic
power spectra as well as realistic initial amplitudes for
perturbations are thus crucial for fully and quantitatively understanding the 
significance of gravitational instability in the formation of giant cloud 
complexes and regulation of global star formation.

\subsection{Program for this work}

In this paper, we investigate both linear and nonlinear evolution of
large scale gravitational instabilities in magnetized, shearing disks
by solving the ideal magnetohydrodynamics (MHD) equations,
and we study the implications of these models for the formation of 
gravitationally bound cloud complexes.
We consider a local patch of an infinitesimally thin gaseous galactic disk 
with physical conditions similar to the solar neighborhood or other 
outer-galaxy regions.  We also consider
models with reduced rotational shear, representing inner-galaxy conditions.
Initially, uniform magnetic fields in the azimuthal direction are included,
and we adopt either an isothermal or an adiabatic equation of state.
We do not include background spiral arm potentials
in the present study, focusing solely on the development of local  
instabilities and their nonlinear outcomes in a featureless gravitational
potential.  Although direct application of our results can be made,
for example, to interarm regions or to flocculent galaxies
where spiral arms are absent or weak, extrapolation of our results with
parameters suitable for the interiors of spiral arms
may yield insight on formation of clouds in 
spiral galaxies with strong arms. Effects of spiral arm potentials 
will be explicitly considered in a subsequent paper.
Related numerical work has been reported very recently by \citet{gam01},
who studied (without magnetic fields) 
the effect of cooling on gravitational instability and
transport of angular momentum by gravitational torques, for application
to accretion disks around young stellar objects or active galactic nuclei.

Our primary objectives in this paper are to understand the 
nonlinear evolution of gravitational instabilities in disks,
and to find the range of $Q$ in high-shear regions that
permits the ultimate formation of gravitationally bound objects.
We also perform a linear analysis with particular emphasis on
the dependence of gravitational instability
upon shear rate and field strength, to investigate how inner- and outer-
galaxy instabilities may differ. 
The linear time evolution of the system we are exploring was
previously studied by \citet{elm87a} and \citet{fan97} (see
also \cite{gam96}), who integrated time-dependent, shearing-sheet 
equations directly, treating the nonaxisymmetric instability as an 
initial-value problem (GLB; JT).
We follow the same ``shearing-sheet'' approach to elucidate the
distinction between two separate instabilities that occur: 
magneto-Jeans instability (MJI)
when shear is weak and/or the magnetic field is very strong; 
and magnetically modified swing amplification (MSA) when shear is relatively
strong and magnetic fields are moderate or weak. 
Because the shear rate varies with radius, we expect the MJI and MSA to 
be important, respectively, in inner and outer galaxies (modulo the effects 
of spiral arms).
MJI and MSA are generalizations, respectively, of the dynamical
processes first studied by \citet{lyn66} and by GLB.
By seeking coherent wavelet solutions, we obtain an algebraic
dispersion relation for magneto-Jeans modes.

Our treatment of gaseous disks as perfectly-conducting, adiabatic or
isothermal monolayers admittedly idealizes the complex, multiphase
interstellar medium (ISM) in real galaxies.
Because we study growth of structure in near-uniform
media with small perturbations, our models are also idealized in
that we consider what would happen to a disk if it were
born in, or evolved to, a state of near homogeneity.
Physically, this initial state could be directly achieved by
the cooling down of a disk that starts hot or has been heated
sufficiently that all large-amplitude perturbations are smoothed
out by pressure forces. Perhaps more practically,
our models also apply to observable galaxies to the extent
that their coherent over-density perturbations on radial scales 
of $\sim3-5$ kpc are not large; as we shall show, smaller-scale
perturbations, although they may initially have significant
amplitude, do not grow strongly. As we shall also show, nonlinear
interactions of small-scale but moderate-amplitude perturbations
may produce larger-scale low-amplitude perturbations, 
which are then susceptible to gravitational runaway if $Q$ is 
sufficiently small. In our simulations, we describe this phenomenon as 
the ``rejuvenated swing''  secondary instability;  
but for real galaxies this might be the most
common way in which gravitationally-bound clouds could grow 
outside of spiral arms.

The organization of this paper is as follows. In the next section,
we present the vertically integrated, 2D MHD equations, and 
describe the characterization of our models via physical and dimensionless
parameters. In \S 3, we revisit previous linear 
analysis to provide coherent wavelet dispersion relations and 
compute total amplification magnitudes,
and thus to distinguish magneto-Jeans modes from  magnetically modified swing
amplifications. We then turn to nonlinear investigations using numerical
MHD methods. The computational code we use and the test results for 
the code performance are described in \S4. In \S5 and \S6, respectively, we 
present the results of local disk simulations with weakly- and 
strongly-shearing background flows. These represent nonlinear evolution
of the MJI and MSA regimes, respectively. We analyze the dependence
of our results on a variety of input parameters, in particular
focusing on what may determine the threshold for collapse, and
what routes the collapse may follow.
We summarize our results and discuss the astronomical implications of our 
work in \S 7.

The reader interested primarily in our quantitative findings for the 
separate gravitational runaway threshold criteria that apply under weak 
shear (inner galaxy) and strong shear (outer galaxy) conditions may wish to
omit the technical sections 2-6 and turn directly to \S 7.

\section{Basic equations and model parameters}
\subsection{Two dimensional magnetohydrodynamic equations}

In this paper we study both linear and nonlinear instabilities of 
galactic disks employing a local, thin-disk approximation.
We consider a patch of the disk whose center lies at a distance $R_0$ from
the galactic center and rotates with a constant angular velocity 
$\Omega(R_0)$ about the galactic center, and work in a local 
Cartesian reference frame with $x$ and $y$ representing
the radial and the azimuthal directions, respectively (GLB; JT). 
We expand the compressible, ideal MHD
equations in the local frame, and neglect terms arising from the 
curvilinear geometry. The equilibrium profile of angular
velocity in the background flow relative to the center of the box at
$x=y=0$ is then expressed by $\mathbf{v}_0 \equiv -q\Omega x \mathbf{\hat{y}}$,
where $q \equiv -d\ln\Omega/d\ln R$ measures the shear.
We finally integrate the resulting equations in the vertical direction 
to obtain the following set of 2D equations:
\begin{equation}\label{con}
  \frac{\partial\Sigma}{\partial t} + \nabla\cdot(\Sigma \mathbf{v}) = 0,
\end{equation}
\begin{equation}\label{mom}
  \frac{\partial\mathbf{v}}{\partial t} +
        \mathbf{v}\cdot\nabla\mathbf{v} = -\frac{1}{\Sigma}\nabla\Pi
       + \frac{1}{4\pi\Sigma}(\nabla\times\mathbf{B})\times\mathbf{B}
       + 2q\Omega^2 x \mathbf{\hat{x}} - 2\mathbf{\Omega}\times\mathbf{v}
       - \nabla\Phi,
\end{equation}
\begin{equation}\label{eng}
  \frac{\partial U}{\partial t} + \nabla\cdot(U\mathbf{v}) = 
       - \Pi\nabla\cdot\mathbf{v},
\end{equation}
\begin{equation}\label{ind}
  \frac{\partial \mathbf{B}}{\partial t} = 
       \nabla\times(\mathbf{v}\times\mathbf{B}),
\end{equation}
and
\begin{equation}\label{Pos}
   \nabla^2\Phi = 4\pi G \Sigma\delta(z),
\end{equation}
(cf, \cite{haw95, gam01}). 
Here, $\Sigma$ is the surface density, $\mathbf{v}$ is the 
vertically-averaged velocity, $\Pi$ and $U$ are the 2D vertically-integrated 
pressure and internal energy, $\Phi$ is the self-gravitational potential, 
and $\mathbf{B}$ is the midplane value of the 3D magnetic field times the 
square root of the unperturbed ratio of surface density to midplane 
volume density.  In equations (\ref{mom}) and (\ref{ind}), we treat the 
effective scale height of the magnetic field as a constant in space and 
time.\footnote{Strictly speaking, the scale height would vary in response to
thermal pressure and magnetic pressure and tension variations.  Our 
simplified treatment neglects force terms arising from the compression and
dilution of magnetic fields by this vertical contraction/expansion, and also
neglects vertical magnetic tension force terms that would arise from scale 
height variations along a given field line.  A full 3D treatment that 
allowed for these terms would permit a study of the coupling between 
the Parker instability and MJI/MSA;  here, by neglecting magnetic scale 
height variations we focus exclusively on the isolating the effects of the 
latter.}
In equation (\ref{Pos}), 
$G$ and $\delta$ are the gravitational constant 
and the Kronecker delta, respectively.

As expressed by equation (\ref{eng}), in this paper we do not include 
thermal heating and cooling effects other than those due to the volumetric 
change (including shocks). We assume an ideal gas equation of state
\begin{equation}\label{EOS}
\Pi = (\gamma-1)U,
\end{equation}
where $\gamma$ is the 2D adiabatic index which differs from
a 3D adiabatic index $\gamma_a$. With $\mathbf{B}=0$, 
in the low frequency limit,
$\gamma = (3\gamma_a-1)/(\gamma_a+1)$ \citep{ost92}, 
while $\gamma = 3 - 2/\gamma_a$ for a strongly self-gravitating disk
\citep{gam01}. In an isothermal medium, $\gamma=\gamma_a=1$.
When $B_{\rm 3D}^2/\rho$ is vertically constant and self-gravity is
not strong, the presence of magnetic pressure yields
$\gamma=(3\gamma_a-1+(2\gamma_a+1)/2\beta)/(\gamma_a+1+3/2\beta)$ for
axisymmetric modes.
In most of simulations presented in this paper, we adopt $\gamma=1.5$, 
corresponding to $\gamma_a=5/3$ for unmagnetized low-frequency perturbations
(for axisymmetric modes with $\gamma_a=5/3$ and $\beta=1$ or $10$, 
$\gamma=1.48$ or $1.50$, respectively). 
As we shall show later, however, instability criteria turn out to be
rather insensitive to the particular choice of $\gamma$.

\subsection{Model parameters}

We consider as an unperturbed initial equilibrium state a homogeneous 
medium having uniform surface density $\Sigma_0$, uniform azimuthal 
magnetic fields $\mathbf{B}_0= B_0 \mathbf{\hat{y}}$, 
thermal sound speed $\cs\equiv(\gamma\Pi_0/\Sigma_0)^{1/2}$, and 
shear velocity profile $\mathbf{v}_0$. We follow the dynamical evolution
of a square domain of size $L$. 
The fundamental dimensional units for length, time, and mass are 
the box edge $L$, the rotation time $\Omega^{-1}$
of the background flow, and the total mass $M_{\rm tot}=L^2\Sigma_0$
contained in the box.
The natural scaling for the other variables is
$L\Omega$ for $\mathbf{v}$, $L^2\Omega^2\Sigma_0$ for $\Pi$ and $U$,
$L\Omega(4\pi\Sigma_0)^{1/2}$ 
for $\mathbf{B}$, and $L^2\Omega^2$ for $\Phi$.
The governing equations (\ref{con})-(\ref{EOS}) in dimensionless form
then depend only on the dimensionless gravitational constant
$g\equiv G\Sigma_0/L\Omega^2$.
In terms of these non-dimensional variables, the initial equilibrium
is represented by $\Sigma=1$, $\mathbf{v}=-qx\mathbf{\hat{y}}$, 
$U=a^2/(\gamma(\gamma-1))$, and $\mathbf{B}=a\beta^{-1/2}\mathbf{\hat{y}}$,
where the dimensionless sound speed $a$ and the dimensionless plasma
parameter $\beta$ are defined respectively by $a\equiv \cs/L\Omega$ and
$\beta\equiv \cs^2/\vA^2$ along with the Alfv\'en  speed $\vA^2
\equiv B_0^2/4\pi\Sigma_0\equiv B_{0,\, 3D}^2/4\pi\rho_0$. 
Therefore, the specification of $a$, $g$, 
and $\beta$ would (together with $q$ and $\gamma$) completely
describe the initial unperturbed configuration of a model disk.

Instead of using $a$ and $g$ directly, we employ two 
input parameters that are equivalent to $a$ and $g$ but 
more illuminating in the context of 
galactic dynamics. One is Toomre's parameter (modified for a razor-thin 
gaseous disk)
\begin{equation}\label{TQ}
Q \equiv \frac{\kappa\cs}{\pi G\Sigma_0} 
        = \frac{a}{\pi g} \sqrt{4-2q},
\end{equation}
where $\kappa$ stands for the epicycle frequency, 
$\kappa^2\equiv R^{-3}d(R^4\Omega^2)/dR$. The other is the Jeans number
\begin{equation}\label{nJ}
n_J \equiv \frac{G\Sigma_0L}{\cs^2} = \frac{g}{a^2},
\end{equation}
of a patch of a 2D thin disk.
For given values of $\cs$ and $\kappa$, the $Q$ parameter measures the
surface density relative to the threshold value at $Q=1$ for axisymmetric 
gravitational instabilities 
(\cite{too64}; see e.g., \cite{bin87, shu92}), 
while $n_J$ is the ratio of the simulation box size $L$ to the shortest 
wavelength $\lambda_J\equiv \cs^2/G\Sigma_0$ permitting
gravitational instability in a thin
(nonrotating) disk. We may define the 2D Jeans mass $M_J$ via
$n_J$ as $M_J\equiv M_{\rm tot}/n_J^2 = \cs^4/G^2\Sigma_0$. 

Our initial disks are smoothed versions of galactic gaseous components,
so it is useful to show how our dimensionless simulation variables 
relate to the large-scale dimensional properties of the Milky Way's
ISM. The atomic plus molecular disks inside the solar circle 
together contribute $7-15\,\Msun\,\pc^{-2}$ (allowing
for He) to the surface density \citep{dam93}. In the solar
neighborhood, the warm components contribute a total surface
density $\sim 6\,\Msun\,\pc^{-2}$ \citep{kul87}, so that the total
solar-neighborhood value is $\sim 13\,\Msun\,\pc^{-2}$. In the
solar neighborhood, the epicyclic frequency is $\kappa=36\,
{\rm km\,s^{-1}\,kpc^{-1}}$ \citep{bin87}, and for a near-flat
rotation curve (corresponding to $q\thickapprox1$) $\kappa\propto R^{-1}$.
Many uncertainties surround both the theoretical concept and 
observational measures of a ``mean'' thermal pressure, but a
number of arguments support a consistent picture with
$P/k\sim 2000-4000\,{\rm K\,cm^{-3}}$ \citep{hei00}. 
With mean midplane density $\sim0.6\,{\rm cm^{-3}}$ \citep{dic90},
this implies a sound speed $\cs = 6-8\,{\rm km\,s^{-1}}$. 
With a mean 3D magnetic field strength 
$B_{0,\, 3D} = 1.4\,\mu\,{\rm G}$ \citep{ran94},
the implied Alfv\'en speed is 
$\vA=3.1 {\rm\,km\,s^{-1}}\,(B_{0,\, 3D}/1.4 \mu\,{\rm G})
(n_{\rm H}/1 {\rm\,cm^{-3}})^{-1/2}$.
We write our dimensionless simulation parameters relative to 
dimensional ISM values as 
\begin{equation}\label{TQn}
Q = 1.4
        \left(\frac{\cs}{7.0 {\rm\,km\,s^{-1}}} \right)
        \left(\frac{\kappa}{36\,{\rm km\,s^{-1}\,kpc^{-1}}}\right)
        \left(\frac{\Sigma_0}{13\,\Msun\pc^{-2}}\right)^{-1},
\end{equation}
\begin{equation}\label{Ln}
L = n_J\times 0.87 \,{\rm kpc} 
      \left(\frac{\cs}{7.0 {\rm\,km\,s^{-1}}} \right)^2
      \left(\frac{\Sigma_0}{13\,\Msun\pc^{-2}}\right)^{-1},
\end{equation}
\begin{equation}\label{beta}
\beta = 6 \left(\frac{P/k}{3000 \,{\rm K\,cm^{-3}}}\right)
          \left(\frac{B_{0,\, 3D}}{1.4\, \mu\,{\rm G}}\right)^{-2},
\end{equation}
for $\gamma_a=5/3$, with the 2D Jeans mass $\Sigma_0\lambda_J^2$ given by
\begin{equation}\label{Jmass}
M_J = 10^7 \Msun \left(\frac{\cs}{7.0 {\rm\,km\,s^{-1}}} \right)^4
  \left(\frac{\Sigma_0}{13\,\Msun\pc^{-2}}\right)^{-1}.
\end{equation}
  From equations (\ref{TQn})-(\ref{Jmass}), therefore, we can see that
with solar-neighborhood values,
our model disk patch with $n_J=5$ corresponding to $L$= 4.4 kpc contains
about $2.5\times 10^8 \Msun$ and is locally stable to axisymmetric 
instabilities. 

Before finishing this section, we remark on a few dynamical time
scales of note. They are the rotation time that we choose as
the time unit in our simulations,
$t_r \equiv 1/\Omega = 3.8\times 10^7\,{\rm yrs}(\Omega
/26 \,{\rm km\,s^{-1}\,kpc^{-1}})^{-1}$, corresponding to
the orbital period,
\begin{equation}\label{Torb}
\torb \equiv 2\pi t_r = \frac{2\pi}{\Omega}
   = 2.4\times 10^8\,{\rm yrs}
     \left(\frac{\Omega}{26 \,{\rm km\,s^{-1}\,kpc^{-1}}}\right)^{-1},
\end{equation}
the sound crossing time
\begin{equation}\label{Tsound}
t_s \equiv \frac{L}{\cs} = 6.3\times 10^8\,{\rm yrs}
    \left(\frac{L}{4.3\,{\rm kpc}}\right)
    \left(\frac{\cs}{7 {\rm\,km\,s^{-1}}}\right)^{-1},
\end{equation}
and a characteristic gravitational contraction time
\begin{equation}\label{Tgrav}
t_g \equiv \frac{t_s}{n_J} = \frac{\cs}{G\Sigma_0}= 
   1.2 \times 10^8\,{\rm yrs}
    \left(\frac{\cs}{7 {\rm\,km\,s^{-1}}}\right)
   \left(\frac{\Sigma_0}{13\Msun\pc^{-2}}\right)^{-1}.
\end{equation}
The growth rate for the fastest-growing ($\lambda=\lambda_J/2$)
gravitationally-unstable mode in a nonrotating disk is $\pi t_g^{-1}$.
The shearing time, corresponding to the time for points on opposite
sides of the box to separate by distance $L$ in azimuth,
is $t_{sh}\equiv 1/q\Omega$, which equals $t_r$ for a disk with
a flat rotation curve ($q=1$).
The small value of $t_{sh}$ compared with $t_s$ and $t_g$
shows that incorporation of galactic differential rotation is essential
in the study of the dynamical evolution of the ISM
on a large scale.

\section{Linear analysis}

Although the linear theory for gravitational instability
in disks with shearing background flows can be found
in GLB for an unmagnetized system, and in \citet{elm87a, elm94}, 
\citet{gam96}, and \citet{fan97} for a magnetized system,
we revisit it with particular attention to the dependence of the 
instability on the shear parameter $q$ and on the plasma 
parameter $\beta$. Our objectives in this section are to obtain
an algebraic dispersion relation for the instability in the limit of 
weak shear or strong magnetic fields (MJI), and to distinguish it from 
the strong-shear instability with relatively weak magnetic fields (MSA). 

\subsection{Linearized equations}

We begin by considering a thin self-gravitating gaseous disk with 
differential rotation, and initially uniform surface density and
uniform azimuthal magnetic fields. We adopt a local 
approximation to investigate the behavior of disturbances whose wavelengths
are small compared with the size of the Galaxy, using the governing set
of equations (\ref{con})-(\ref{EOS}).
The stability of a local patch of the disk
can be studied in the shearing sheet coordinates 
$(x',y',t') = (x,y+q\Omega xt, t)$ (GLB; JT).
We consider the time development of
an initial plane-wave disturbance which preserves sinusoidal
variations in the local rest frame of the equilibrium
shearing, rotating flow, i.e.
\begin{equation}\label{chi}
\chi_1(x',y',t') = \chi_1(t')\exp(i\kx x' + i\ky y'),
\end{equation}
where $\chi_1$ refers to any perturbed physical variable and
$\kx$ and $\ky$ represent the initial respective wavenumbers along the
$\mathbf{\hat{x}}$- and $\mathbf{\hat{y}}$-directions. 
We linearize the MHD equations 
(\ref{con})-(\ref{EOS}) and apply perturbations of the form 
(\ref{chi}) in the transformed coordinates. 
The resulting equations can be written
(omitting the prime on $t$) 
\begin{equation}\label{lcon}
\frac{d\sigma}{dt} = -\kx(t) u_x - \ky u_y,
\end{equation}
\begin{equation}\label{lmomx}
\frac{du_x}{dt} = 2\Omega u_y 
  + \kx(t)\left(\cs^2-\frac{2\pi G\Sigma_0}{|k(t)|}\right)\sigma
  - \vA^2k(t)^2 m,
\end{equation}
\begin{equation}\label{lmomy}
\frac{du_y}{dt} = -\frac{\kappa^2}{2\Omega}u_x 
  + \ky \left(\cs^2-\frac{2\pi G\Sigma_0}{|k(t)|}\right)\sigma,
\end{equation}
\begin{equation}\label{lmag}
\frac{dm}{dt} = u_x.
\end{equation}
Here, $\sigma\equiv\Sigma_1/\Sigma_0$, $\mathbf{u}\equiv i\mathbf{v}_1$, 
the dimensionless vector potential $m$ is defined through
$\mathbf{B}_1 \equiv -i B_0\nabla\times(m\mathbf{\hat{z}})$, 
the time dependent wavenumber 
$\kx(t) \equiv \kx + q\Omega\ky t$, and $k(t)^2\equiv
\kx(t)^2 + \ky^2$.

Equations (\ref{lcon})-(\ref{lmag}) are ordinary differential 
equations in time and the explicit time-dependence of $k(t)$ does
not permit eigensolutions in general. This is the characteristic
of any system with background shear. An applied spatial planform 
is wrapped up from leading to trailing configuration by the kinematics
of shear; its radial wavenumber increases linearly with time. The
response of the system to nonaxisymmetric disturbances can be studied
through direct numerical integrations of the linearized equations
as an initial value problem.
\citet{elm87a} followed this approach to investigate nonaxisymmetric
gravitational instabilities in the magnetized gaseous galactic disk
as a supercloud formation mechanism. 
\citet{gam96} computed the nonaxisymmetric responsiveness of 
a magnetized disk under large differential rotation to gravitational
instabilities, and showed that magnetic fields reduce the responsiveness.
\citet{fan97} extended Elmegreen's work to study the long-term
evolution of MHD density waves. We refer the reader to
these works for illustrative examples of perturbed
density evolutions.

\subsection{Coherent wavelet analysis}

While the general response of the system to nonaxisymmetric 
perturbations is found from the temporal integrations of 
equations (\ref{lcon})-(\ref{lmag}),
there are certain regimes in the parameter space in which we 
can seek time-localized ``coherent wavelet'' solutions having
the same time dependence for all perturbed variables. The condition
for the existence of the coherent solutions is that the time
over which the instantaneous growth rate 
\begin{equation}\label{zeta1}
\zeta(t) \equiv \frac{d \ln\chi_1(t)}{dt}, 
\end{equation}
changes is sufficiently large compared with
the growth time $\zeta^{-1}$, that is,
\begin{equation}\label{zeta2}
\left|\frac{d\ln \zeta(t)}{dt}\right| \ll \zeta(t)
\end{equation}
\citep{kim00}. This amounts to a temporal WKB limit. Since 
equation (\ref{zeta2}) can be written as 
$(q\Omega/\zeta)(\kx(t)\ky/k^2)(d\ln\zeta/d\ln k) \ll 1$,
the existence of coherent wavelet solutions is guaranteed 
if shear is weak $(q\Omega \ll \zeta)$, 
if the instantaneous growth rate is insensitive 
to $k$ ($d\ln\zeta/d\ln k \ll 1)$, or
if modes are near-axisymmetric $(\kx(t)\ky/k^2 \ll 1)$. This technique
has been applied to study the effect of differential rotation 
on the Parker instability in the Galactic disk \citep{shu74} and 
on the convective instability in accretion disks \citep{ryu92}.
Very recently \citet{kim00} derived an analytic dispersion
relation for nonaxisymmetric magnetorotational instabilities via
the coherent wavelet formalism in a strong shear environment.

The coherent wavelet solutions for magneto-Jeans instabilities
are immediately obtained by substituting equation (\ref{zeta1})
into equations (\ref{lcon})-(\ref{lmag}) and then putting
$d\chi_1/dt \rightarrow \zeta\chi_1$ following the
approximation (\ref{zeta2}). The nontrivial solutions of the 
resulting equations obey a quadratic equation in $\zeta^2$:
\begin{equation}\label{sol4}
\zeta^4 + [\kappa^2 - 2\pi G \Sigma_0 |k|
            + (\cs^2+\vA^2)k^2]\zeta^2
 + [\cs^2k^2 - 2\pi G \Sigma_0 |k|] \vA^2\ky^2 = 0.
\end{equation}
Note that the time-dependence of $\zeta$ is absorbed implicitly 
in $k=k(t)$.
It can be shown that the coherent solution (\ref{sol4}) is 
self-consistent if $q \ll 1$ (a weak-shear limit)
or if $\cs^2 \ll \vA^2$ (a strong-field limit). Therefore,
when $\vA^2\ky^2 \neq 0$, any modes that satisfy
the instantaneous instability criterion 
\begin{equation}\label{kJ}
k(t) < k_J \equiv \frac{2\pi G\Sigma_0}{\cs^2},
\end{equation}
are subject to a transient or an exponential growth. 
For fixed $|k|$, $\zeta$ is maximized with $\kx=0$.

Two special cases of equation (\ref{sol4}) deserve some comment.
When $\ky=0$, corresponding to axisymmetric modes, the last term
in equation (\ref{sol4}) vanishes and $k$ becomes a constant. In 
this case, equation (\ref{sol4}) is reduced to the dispersion
relation for density waves with a stabilizing role played 
by magnetic fields \citep{gam96, lou98}. 
With $\vA=0$, one can show that a disk becomes unstable only if
$Q<1$, and the range of unstable wavenumbers is
$1-(1-Q^2)^{1/2} < \kx/k_{\rm max} < 1 + (1-Q^2)^{1/2}$,
where $k_{\rm max}\equiv k_J/2=\pi G\Sigma_0/\cs^2$.
The maximum growth rate is $\zeta_{\rm max} = \cs k_{\rm max}
(1-Q^2)^{1/2} = \kappa(Q^{-2}-1)^{1/2}$ 
at $\kx=k_{\rm max}$ (e.g., \cite{shu92}).  
Presence of magnetic fields modifies the Toomre criterion
for a hydrodynamic disk in such a way that magnetized disks 
are unstable to axisymmetric perturbations if
$Q_M \equiv \kappa(\cs^2+\vA^2)^{1/2}/\pi G \Sigma_0 =
Q(1+1/\beta)^{1/2} < 1$; the maximum unstable wavenumber in a nonrotating 
disk is 
given by equation (\ref{kJ}) with $\cs^2\rightarrow(\cs^2+\vA^2)$,
and the fastest growing wavenumber in a rotating disk is half of this. 
The enhanced axisymmetric stability associated with 
magnetic fields changes dramatically when nonaxisymmetric modes
are considered.

The other interesting limit of equation (\ref{sol4}) 
is rigidly rotating disks with $q=0$. Without shear, $\mathbf{k}$ 
is again
independent of time, and thus equation (\ref{sol4}) is the
exact solution to  equations (\ref{lcon})-(\ref{lmag}).
When $q=0$, the equilibrium may contain nonzero radial magnetic 
fields, so we may generalize equation (\ref{sol4}) by
replacing $\vA\ky$ with $\mathbf{\vA}\cdot\mathbf{k}$ in the last term.
Magnetic destabilization is apparent and one can easily
see that the stability criterion is the 2D Jeans condition:
regardless of the magnetic field strength,
magnetized disks in rigid-body rotation are subject to nonaxisymmetric 
gravitational instabilities if the condition (\ref{kJ}) is
satisfied. Magnetic tension from bent field lines can reduce
the stabilizing effect of the Coriolis force by resisting
fluid displacements in the radial direction.
When the field is very strong ($\vA$ large; $\beta \ll 1$), motion of gas is
mainly parallel to the field lines,
so perturbations evolve as if there were no rotation, yielding
the same dispersion relation 
$\zeta^2=(2\pi G \Sigma_0 |k|-\cs^2k^2)(\mathbf{\vA}\cdot\mathbf{k})^2
/\vA^2k^2$ as in a nonrotating, thin disk. With a weak magnetic field 
($\beta \gg 1$), the fluid motion is not strictly parallel to
field lines and thus growth rates (when $Q>1$) become small 
as $\zeta\propto\beta^{-1/2}$.
This destabilizing effect
of magnetic fields was first studied by \citet{cha54} 
for an infinite homogeneous medium and by \citet{lyn66}
for a thin galactic disk.\footnote{Although  equation (\ref{sol4})
has the same form as equation (1) of \citet{lyn66},
equation (\ref{sol4}) is more general in the sense that 
it allows for the solutions with $q\neq 0$ via the 
value of $\kappa^2=(4-2q)\Omega^2$ and the time dependence of 
$\kx(t)=\kx+q\Omega\ky t$.}
The detailed physical description is given by \citet{elm87a}.

\subsection{Magneto-Jeans instability {\it vs.} the magnetically \\
modified swing amplifier}

In the general case of nonaxisymmetric perturbations with nonvanishing
shear, the secular increase of $k(t)$ would inevitably suppress 
the development of the instability, implying a transient
growth of perturbations: the system is instantaneously unstable
only when $|t| < t_c \equiv (K_y^{-2}-1)^{1/2}/q\Omega$, where
$K_y\equiv \ky/k_J$. In order
to quantify the virulence of the instability, we define an
amplification magnitude as
\begin{equation}\label{amp-mag}
\Gamma \equiv \log \frac{\Sigma_1(t_c)}{\Sigma_1(-t_c)}
     = 2 (\log e) \int_0^{t_c} \zeta(t) dt.
\end{equation}

Amplification magnitudes for coherent wavelet solutions are easily 
computed from equations (\ref{sol4}) and (\ref{amp-mag}).
The resulting $\Gamma$ for $Q=1.3$ and $K_y=0.5$ is displayed
in Figure \ref{fig-qGam} with dotted lines. Also shown 
as solid lines are the results of direct numerical 
integrations of the shearing sheet equations 
(\ref{lcon})-(\ref{lmag}). The excellent agreement for
small $q$ and/or for small $\beta$ between the
results from the two different approaches demonstrates the validity
of the coherent wavelet analysis for the weak-shear or strong-field limit.
For $Q\geq 1$ and in the weak-shear limit,
the solutions of equations (\ref{sol4}) and (\ref{amp-mag}) 
corresponding to the fastest growing modes that 
have $K_y=1/2$ for $\beta\ll 1$
and $K_y\sim 1/2 - 3/4$ for $\beta\simgt 0.1$
can be approximated as
\begin{displaymath}
\nonumber
\;\;\;\;\;\;\;\;\;
\;\;\;\;\;\;\;\;\;
\;\;\;\;\;\;\;\;\;\;
\Gamma_{\rm max} = \left\{ \begin{array}{ll}
\frac{2\sqrt{1-q/2}}{qQ}, 
&{\rm for\;\;\;} \beta\ll 1 \;\;{\rm and}\;\; q\simlt 1, 
\;\;\;\;\;\;\;\;\;\;\;\;\;\;\;\;\;\;\;\;\;\;
\;\;\;\;\;\;\;\;\;\;\;\;\;\;\;\;\;\;\;\;
{\rm (26a)}
\\
\\
\frac{1.6}{q\sqrt{\beta}}\frac{1}{Q^2-f(\beta)},
&{\rm for\;\;\;}\beta\simgt 0.1 \;\;{\rm and}
\;\; q\simlt 0.7\beta^{-1/3}Q^{-1},
\;\;\;\;\;\;\;\;\;\;\:
\;\;\;\;\;\;\;\;\;\;\:
{\rm (26b)}
\end{array}\right.
\end{displaymath}
\setcounter{equation}{26}
where $f(\beta)$ is defined by 
\begin{equation}\label{fbeta}
f(\beta) \equiv 1.17 - \frac{3.97}{(\log{\beta}+2.16)^{1.48}}.
\end{equation}
For $10^3>\beta\geq 1$, $|f(\beta)|<1$, so that $f(\beta)\simeq 0$ is 
a good approximation for $Q>2$.
$\Gamma_{\rm max}$ from equations (26) and
(\ref{fbeta}) are within 
$\sim 10\%$ of the results of direct shearing sheet
integrations.
As both equations (26) and Figure \ref{fig-qGam}
show, $\Gamma\propto q^{-1}$ for $q\ll 1$, which
is a manifestation of the longer time interval for the growth of $\kx$;
in equation (\ref{sol4}), the instantaneous growth rate 
$\zeta$ is very weakly dependent on $q$
only through $\kappa^2=(4-2q)\Omega^2$. 
Equation (26a) gives the maximum amplification 
amplitude for all $\beta$.

One interesting feature in Figure \ref{fig-qGam} is 
the $\beta$-dependence of $\Gamma$ obtained from the temporal integration
of the shearing sheet equations. When $\beta \ll 1 $, $\Gamma$ is a
monotonic function of $q$, well approximated by the coherent solutions.
When $\beta \gg 1$, however, $\Gamma$ first decreases exponentially with 
increasing $q$ from zero, becoming vanishingly small at intermediate
$q$, and rapidly rises again to show local maxima at $q\sim 1$.
With $q\ll 1$ and $\beta \gg 1$, $\Gamma \propto
\beta^{-1/2}$ (see eqs.\ [26]). 
In the extreme unmagnetized limit $\beta\rightarrow \infty$, 
the system is completely stable for $q\ll 1$, while showing
moderate growth ($\Gamma \simlt 1.6$) when $0.2<q<2$. Evidently,
there are two different types of instabilities. The first kind,
operating very efficiently when $\beta \ll 1$, {\it or}
$q\ll 1$ and $\vA\neq0$, 
is similar to the Jeans instability in nonrotating disks.
We refer to this as the magneto-Jeans instability (MJI).
In the second type of instability,
operating when $\beta \simgt 1$ {\it and} $q\sim 1$, disturbances grow
from the swing amplifier mechanism (GLB; \cite{too81}) and 
magnetic fields are not required; in fact, magnetic fields reduce 
the amplification factor. We refer to this swing-related instability 
as magnetically-modified swing amplification (MSA). 

The contrasting role played by magnetic fields in MJI and MSA is well
illustrated in Figure \ref{fig-Qbeta} where we plot $\Gamma$ with contours
on a $Q-\beta$ plane. $K_y=0.2$ and $q=1$ are adopted and only the results
from temporal integrations of equations (\ref{lcon})-(\ref{lmag}) are
presented. Notice the discontinuity in $\Gamma$ that clearly 
separates the range of $\beta$ into two parts. Defining $\beta_c$ as 
the value of $\beta$ at the discontinuity, one can see
that the MJI dominates for $\beta < \beta_c$, while a system with 
$\beta>\beta_c$ is more susceptible to MSA.
For a range of the parameters $K_y$ and $Q$, $\beta_c \sim  0.1 - 1$.
For a given $Q$, $\Gamma$ attains its minimum at $\beta=\beta_c$. 
When $\beta > 30$,
$\Gamma$ is almost independent of $\beta$, indicating that modification
of swing amplification by magnetic fields becomes negligibly small.
When $q=1$, MJI is less sensitive to $Q$ than MSA is.

As clearly explained by \citet{too81} and \citet{bal88}, hydrodynamic
swing amplification arises as a consequence of the conspiracy
between three agents: background shear, epicyclic shaking, and 
self-gravity. Since wave fronts sweep in the same sense as epicycle motions
from a leading to a trailing configuration, fluid elements in the wave crests 
remain longer in the region of excess density, extending their exposure
to self-gravity. The net effect is significant growth of the wave
amplitude. The fact that swing amplification depends essentially
on shear, expressed by the time-dependence of $\kx(t)$, makes
MSA different from MJI, in which shear has only a 
stabilizing effect. This explains why solutions for MSA cannot be
found in the coherent regime that treats $\kx(t)$
as a constant instantaneously\footnote{\citet{bin87} also noted
that the spatial WKB approximation fails to capture the swing 
amplification.}.
Since rotation curves of spiral galaxies are generally flat
(see, e.g., \cite{sof99}) with $q\sim 1$ except in small regions very close to
their centers, and given the likely condition of near- or sub-
equipartition of magnetic energy relative to 
thermal energy ($\beta\sim 1-10$), MSA is expected to be more important 
than MJI in the large-scale dynamical evolution of disk galaxies
with weak spiral structure.
On the other hand, in the central regions of spiral
galaxies where rotation curves are almost linear, or
in irregular galaxies where shear is relatively weak, MJI 
may play a very important role in the formation of molecular clouds.

Previously, these MJI and MSA modes have been referred to as the
swing amplifications of slow
and fast MHD density-wave modes, respectively, by \citet{fan97}.
We adopt our nomenclature because it more transparently refers to
the underlying physical mechanisms.
Even though slow MHD waves become unstable if
the condition (\ref{kJ}) is satisfied, it is not precisely 
a ``swing'' amplification  but more like the 2D Jeans instability with
magnetic fields resisting the stabilizing Coriolis force. Moreover,
when they grow, MSAs do not maintain as clear wave 
properties as the fast MHD waves do, as the nonexistence of 
coherent wavelet solutions for MSAs suggests.

Being isotropic, acoustic waves (or random motions of stars in a 
stellar disk) affect swing amplification simply by reducing 
the equivalent gravitational force (JT; \cite{too81}), 
and it is sound waves that
eventually stop the growth of hydrodynamic swing amplifier.
One may ask whether the role of magnetic fields represented by 
the Alfv\'en speed $\vA$ is the same as that of $\cs$ in MSA, as is true 
in the axisymmetric case for MJI (\S3.2). The answer is that
magnetic effects instead act as an ``anisotropic'' reduction factor for
self-gravity. The stabilizing effect from magnetic fields 
is via magnetic pressure. With initial azimuthal fields, however,
the magnetic pressure force in the linearized form has only
a radial component and requires $\kx\neq 0$. Therefore,
stabilization of MSA due to field lines 
occurs mostly at intermediate pitch angles of the wave fronts,
and is vanishingly small when $\kx\sim 0$, while
thermal pressure still stabilizes a system even when $\kx=0$.

In Figure \ref{fig-KGam}, we show the effect of varying $K_y$ on MSA.
We take $Q=1.1, 1.5$ and $q=1$. Near $K_y\sim 1$, 
thermal and magnetic pressures cause $\Gamma$ 
to decrease rapidly for all $\beta$. 
At the opposite end of the curves, the 
reduction of $\Gamma$ with $\beta\rightarrow\infty$ is due
to the enhanced relative importance of the epicyclic oscillation
that is independent of $K_y$. 
For MJI modes with $\beta \ll 1$, the stabilizing effect 
from the epicyclic motion appears at small $K_y\sim 0.1\beta$. 
The discontinuities in $\Gamma$ on the $\beta=1$ curves again
emphasize the existence of two different instabilities:
MSA operates with small $K_y$, while strong tension forces
with large $K_y$ favor MJI.

A recurring issue in studying dynamics of massive disks is the question
of how large a scale must be considered, due to the nonlocal
nature of gravity. \citet{bal99}, for example, suggest that 
angular momentum transport by gravitational torques should be studied in the 
framework of a global rather than local model, although they point out 
that this is less of an issue for disks with $Q$ near unity,
because wave propagation is less important. Note that for $\beta\geq 1$, 
we find (see e.g., Fig.\ \ref{fig-KGam})
maximum values of $\Gamma$ are achieved at 
$K_y \sim 0.15-0.4$, corresponding to a few kpc from equation (\ref{Ln}). 
This implies that the local model we will
adopt for our simulation study is somewhat marginal as a direct 
representation of a small piece of a galactic disk. In spite of this
caveat, we believe the model still captures the important
features of instabilities growing in self-gravitating,
thin disks, at least as long as $\beta \geq 1 $.
We discuss checks of this obtained by varying the simulation
box size, in \S6.

To summarize this section, we found that shearing, self-gravitating
disks with embedded magnetic fields are unstable to two kinds of
nonaxisymmetric instabilities:
(1) the magneto-Jeans instability, which works efficiently in a 
strongly magnetized medium  or in a medium with 
weak shear, and (2) magnetically-modified
swing amplification, which works in a moderately or 
weakly magnetized medium with 
relatively strong shear. Increased magnetic fields tend to destabilize 
the former, while stabilizing the latter.
We provided coherent wavelet solutions applicable to MJI.

\section{Numerical method and code tests}

To follow the nonlinear evolution of gravitational instabilities
in shearing, magnetized disks, we integrate equations 
(\ref{con})-(\ref{EOS}) using a modified version of the ZEUS code 
for numerical MHD \citep{sto92a, sto92b}. ZEUS uses a time-explicit,
operator-split, finite-difference method to solve the MHD equations
on a staggered mesh.
The MHD algorithm employs the constrained-transport formalism
to maintain the condition $\nabla\cdot\mathbf{B}=0$ within machine
precision, and the method of characteristics for accurate
propagation of Alfv\'en waves \citep{eva88, haw_sto95}.
For the advection of the hydrodynamic variables, we decompose the 
azimuthal direction ($\mathbf{\hat{y}}$) velocity into the mean shearing part 
and the remaining
perturbed part, and transport only the perturbed terms with
ZEUS's advection algorithm, while treating the contribution arising
from the mean shearing part as source terms\footnote{If we decompose
the velocity as $\mathbf{v}=\mathbf{v}_0 + \delta\mathbf{v}$,
the advection term of the $\mathbf{\hat{y}}$-momentum due to 
the perturbed velocity in conservative form appears as
$\nabla\cdot(\Sigma v_0 \delta \mathbf{v}) =
-q\Omega\Sigma\delta v_x + v_0\nabla\cdot(\Sigma\delta\mathbf{v})$,
the first term of which can be evaluated in the source step.
There are various ways to compute $\Sigma\delta v_x$ in 
a finite difference scheme, but we found that
\begin{displaymath}
\Sigma\delta v_x\big|_{i,j} =
\frac{1}{4}\big[\Sigma_{i,j}(\delta v_{x,i+1,j} + \delta v_{x,i,j})
+ \Sigma_{i,j-1}(\delta v_{x,i+1,j-1} + \delta v_{x,i,j-1})
\big],
\end{displaymath}
gives the most accurate results. Here, the indices $i$ and $j$ denote
respectively the $x$- and $y$-coordinates of a spatial staggered grid point
defined in the ZEUS code.}.
Through an advection test of a square pulse moving across the radial
boundaries, we confirmed that the velocity decomposition method gives
less diffusive results than does transport involving the whole velocity,
especially when the mean shearing velocity is comparable to
or larger than the transverse velocity component.
Another numerical method for treating a strongly supersonic shear
flow in a non-self-gravitating, unmagnetized medium is described 
by \citet{mas00}; a similar technique was adopted by \citet{gam01}.

We apply shearing box boundary conditions in which the azimuthal boundaries 
are perfectly periodic and the radial boundaries are shearing-periodic.
In the absence of initial radial magnetic flux, the shearing box boundary 
conditions conserve the total magnetic flux through each surface of the 
sheared box; the long-time average through the $\mathbf{\hat{y}}$-surfaces 
of the computational box is then also uniform. 
The volume-integrated total energy in a shearing box
is not a conserved quantity; its rate of change is 
determined by the angular momentum flux associated with
Reynolds, Maxwell, and gravitational stresses across the radial
surfaces \citep{haw95, bal99, gam01}.
We implemented the remap algorithm discussed by \citet{haw95} in which
the physical values at the ghost zones adjacent to one radial 
boundary are remapped in the Lagrangian sense from the values 
at the active zones on the other radial boundary.
Because of the opposite background velocities, however,
the flow characteristics at each radial boundary are quite different
from each other. After the remap procedure, therefore, discontinuities
in the flow characteristics are unavoidable across the radial boundaries,
tending to generate vorticity when fluid moves across the radial boundaries
and adding noise to propagating \alf waves.
We address this problem by keeping the ghost zones adjoining the 
radial boundaries active in the transport step along 
the $\mathbf{\hat{y}}$-direction.

The solution of the Poisson equation is obtained by the 
Fourier transform method modified by the shearing box boundary
condition, as discussed by \citet{gam01}. Since the density
distribution at an arbitrary time is not periodic in the 
$\mathbf{\hat{x}}$-direction,
we transform to sheared coordinates, $x'=x, y'=y+q\Omega x (t-t_p)$, 
after finding
the nearest time $t_p$ (either positive or negative) that gives
an exactly periodic distribution in the transformed coordinate.
We solve the Poisson equation using the standard discrete 
fast Fourier transform technique in the primed coordinates and 
then transform the calculated
potential back into the simulation domain. To make the gravitational
force isotropic on small scales, we discard the modes with
$|k|> 2^{-1/2}(N/2)(2\pi/L)$, 
where $N\times N$ is the numerical resolution.

For most of our simulations (except where noted), 
initial perturbations are realized by
a Gaussian random density field with a power spectrum 
$\langle|\Sigma_k|^2\rangle \propto k^{-8/3}$ 
for $2\pi/L \leq k \leq (N/2)(2\pi/L)$.
If the perturbations arise from motions obeying a sonic dispersion
relation ($\omega\sim\cs k$), this corresponds to the 2D Kolmogorov
velocity spectrum that gives a conservative cascade of energy transfer.
This choice of power spectrum is to some extent arbitrary; one may 
argue, for example, whether a Kolmogorov or Burgers power spectrum is 
more appropriate. Through the numerical simulations, however, 
we have confirmed that the particular choice of spectral index  
does not make a large difference, as long as the power spectrum is decreasing
at large $k$. We measure the standard deviation $\epsilon_0$ of the 
initial density fluctuations $\Sigma/\Sigma_0-1$ in real space, 
and use this value to parameterize the initial
perturbation amplitudes; we allow $\epsilon_0$ to vary from
$10^{-3}$ to $10^{-2}$. 

\subsection{Code tests}

We have verified our numerical algorithm with a wide variety of test
problems: MHD shock tubes and propagation of polarized Alfv\'en
waves in nonshearing medium, advection of a square box with
enhanced density in a background shear flow, 
axisymmetric traveling and standing MHD waves
modified by self-gravity and shear, and
axisymmetric gravitational instabilities with shear.
By comparing the results with corresponding analytic predictions,
we have confirmed the code's performance.

Of particular importance for the current work is the test 
calculation of linear,
nonaxisymmetric gravitational instabilities, because this test
can directly verify the correctness of our implementation
of the boundary condition and the Poisson solver.
In order to compare with the results of linear analysis,
we initially apply very small amplitude sinusoidal perturbations,
$\Sigma_1=\epsilon_0\cos(\kx x+\ky y)$, 
$v_{y,1} =\epsilon_0\sin(\kx x+\ky y)$,
and $B_{x,1} = \epsilon_0\ky/2\pi\cos(\kx x+\ky y)$;
the initial conditions for the other variables were determined by
imposing the requirements of adiabatic, zero-vorticity, 
$\nabla\cdot\mathbf{B}_1=0$ perturbations. 
The other chosen parameters are $Q=2$, $n_J=2.5$, $q=1$,
$\gamma=1.5$, $\beta=1$, 
$\epsilon_0=10^{-4}$, $\kx/2\pi=-6$,
and $\ky/2\pi=1$.
In Figure \ref{fig-test} we display with solid line the Eulerian time
evolutions of the variables at a location $(x,y)=(-0.07,0.28)$
from a $128^2$-zone simulation, which are in excellent agreement
with the results from linear analysis plotted by open circles.
During the swing amplification ($t\sim 2-9$) phase, the system experiences
an increase of density by a factor of 25. As time increases 
the MSA phase ends and density oscillates in accordance with the passage
of waves of high $k$. For all the variables,
a small discrepancy between linear and nonlinear solutions is observed
at $t\sim10-12$. This is not an error related to the numerical scheme.
Rather, this is due to modulation of the wave mode of our
interest with waves that are generated from nonlinear effects
and subsequently grow with swing amplification. 

\section{Nonlinear evolution of magneto-Jeans instability}

Following up the linear analysis of MJI in \S3, we have performed 
numerical simulations of the development of these instabilities
in a medium with weak shear,
adopting $Q=1.5$, $n_J=5$, $q=0$ or 0.1, and $\beta=1$, while
varying $\gamma$ from 1 to 2. 
We initially perturbed a uniform density medium by applying 
white noise perturbations\footnote{Since the
fastest growing mode has $\ky=3.3(2\pi/L)$ for the chosen parameters,
power-law perturbations that have most of the power at $\ky\simeq0$
would exhibit complicated behavior in the evolution of 
maximum surface density at early times, as the initial growth of low-$k$ modes
is succeeded by faster-growing but initially lower-amplitude higher-$k$ modes. 
In order to study the modal growth
of perturbations as cleanly as possible, we have instead imposed
white noise perturbations for these MJI simulations. Because
MJIs with $q\ll 1$ yield almost exponential growth,
the final state, always dominated by the most unstable modes,
is quite independent of the type of applied perturbation.} 
of amplitude $10^{-3}\Sigma_0$,
and then evolved the system up to $t/\torb=4$.
The resulting time histories of the maximum density for $q=0.1$ and
varying $\gamma$ are plotted in Figure \ref{fig-MJI-devol}.
For comparison, we also plot with dotted line the result from the simulation 
with $q=0$ and $\gamma=1$,
for which the kinematic stabilization of MJIs is absent.
Exemplary snapshots of density structures, perturbed velocity, 
and magnetic field 
lines at $t/\torb=2, 4$ of the $\gamma=2$ simulation are displayed in 
Figure \ref{fig-MJI-image}.

Initially, all the modes that satisfy the instantaneous 
instability criterion (\ref{kJ}) begin to grow, and subsequently
the system becomes dominated by the most unstable mode. 
With the chosen parameters,
the coherent wavelet solutions predict that the mode with $K_y=0.66$
corresponding to $k_y = 3.3(2\pi/L)$ has the largest instantaneous 
growth rate at $k_x=0$. 
When the system is rigidly rotating with $q=0$,
this mode is expected to continue to grow 
exponentially without limit.
When $q=0.1$, on the other hand, this mode
is allowed to grow only until $t/\torb=3.6$ when the radial 
wavenumber increases 
up to $k_x(t) \simeq 3.8 (2\pi/L)$,
forming corrugated stripes 
with a pitch angle $\phi \equiv \tan^{-1}(\ky/\kx)= 41^{\rm o}$.
Beyond this time, thermal pressure dominates self-gravity
and no further growth is expected.
The estimated total amplification for the $q=0.1$ case is about 8 orders
of magnitude. 

The linear prediction for the most unstable mode 
is indeed shown in the simulations. 
Figure \ref{fig-MJI-devol} shows that for $q=0$,
the density increases exponentially with time
until nonlinear effects control the dynamics. With the
isothermal equation of state, the density rises more rapidly
as the overdense regions start to collapse nonlinearly.
The growth rates of the maximum density for $q=0.1$ are not
much different from the rigidly rotating case, because 
$k_x$ remains small 
throughout the linear phase.
In fact, the maximum pitch angle attained before
the nonlinear effects become significant 
is $\phi\sim 53^{\rm o}$ corresponding to
$k_x=2.5(2\pi/L)$ (see, for example, Figure \ref{fig-MJI-image}), 
indicating that kinematic stabilization from shear never occurs.
The late nonlinear phase (but not the linear phase) of MJI is 
very sensitive to the choice of $\gamma$.
When $\gamma$ is close to 1, overdense stripes collapse rapidly as they 
form mainly along the azimuthal direction.
As $\gamma$ increases, the collapse is progressively delayed so that 
the stripes have more time to be sheared out. 
If the perturbed velocity
field has grown almost comparable to or even greater than
the background shear velocity, the azimuthal shear rate is also
locally modified substantially from the initial value. 
Gravitational attraction between neighboring stripes also changes 
the shear rate significantly, causing them eventually to collide with
each other.
The result is the formation of a few big condensations. 
When $\gamma <1.6$, the condensed lumps keep collapsing towards 
their centers, making it difficult to follow the subsequent evolution. 
We stop the simulation when the central density of the biggest lump
becomes so high that we are not able to spatially resolve the central 
region. In general, in order to prevent spurious results arising
from limited numerical resolution,
the local grid scale must be smaller than the smallest
local gravitationally unstable wavelength \citep{tru97, tru98};
since the instantaneous value of $k_J$ (eq.\ [\ref{kJ}])
increases with increasing $\Sigma$, any unstable low-$\gamma$ simulation
on a fixed grid eventually becomes unresolved.

When $\gamma \simgt 1.6$, on the other hand, a stiff equation of state 
provides the strong thermal pressure gradient to halt gravitational 
collapse, and form gravitationally bound clumps.
The right frame of Figure \ref{fig-MJI-image}, for example,
shows a disk consisting of 4 clumps and ridges
connecting them. The two biggest clumps to the left have
each 32\% and 17\% of the total mass. Additional support
from magnetic pressure makes them elongated across the field lines.
The field strength is correlated with the surface density roughly as
$B\sim \Sigma^{0.5}$. 
Clumps grow slowly as they accrete material and then
experience collisional agglomeration to form
ultimately one large object, after $t/\torb=5.4$.

\section{Nonlinear evolution of magnetically modified swing amplification}

In the previous section, we showed that 
when shear is very low, rotating, magnetized, self-gravitating
systems are unstable to nonaxisymmetric disturbances,
eventually collapsing or forming gravitationally bound condensations
within $\sim$2 orbital times. This result is interesting
in the sense that inclusion of magnetic fields modifies the
dynamics significantly, permitting instability even when $Q>1$, 
possibly accounting for the active star 
formation towards centers of galaxies (cf.\ \cite{ken93, ken98, mar01}).
However, the majority of material in
spiral galaxies in fact orbits subject to a flat rotation curve (at least
away from strong spiral arms).
With such strong shear, MJI is disabled, but MSA becomes active.
To explore nonlinear evolution of such high-shear regions under MSA,
we have carried out extensive simulations by fixing $q=1$ and varying
other input conditions.

\subsection{Hydrodynamic case}

Table \ref{tbl-1} lists the various parameters for simulations 
following the nonlinear evolution of swing amplification
in purely hydrodynamic disks. The first column labels each run with the 
prefix H and a model number. The second and 
third columns give the two basic parameters of our simulations:
Toomre's stability parameter $Q$ and the Jeans number $n_J$ of 
the simulation box
(see eqs.\ [\ref{TQ}] and [\ref{nJ}]).
The 2D adiabatic index for the equation of state is given in 
the fourth column, while the fifth column gives the amplitude
of initial perturbations indicated in terms of 
the standard deviation $\epsilon_0$ of density fluctuations.
The numerical resolution is shown in the sixth column. The next two columns 
list the termination time $\tau_{\rm 1st}$ of the MSA and the
amplification magnitude $\Gamma_{\rm 1st}$ that measures the 
growth in the standard deviation of the density distribution
at $t=\tau_{\rm 1st}$ determined from the simulation results.
The time $\tau_{\rm 1st}$ is defined as the time corresponding
to the first local maximum in the evolutionary history of
the RMS density fluctuation, $\epsilon \equiv
\langle(\Sigma/\Sigma_0-1)^2\rangle ^{1/2}$.
Times are given in units of $\torb$ (see eq.\ [\ref{Torb}]).
The ninth column gives the time when 
the nonlinear secondary instability (see below) begins.
The final simulation outcome (i.e.\ within $4\,\torb$) is given in 
the final column where ``Unstable'' means a runaway collapse or 
the formation of condensation, while ``Stable'' means 
low-amplitude wave motion at the end of the simulation. Cases that do not
produce any condensations but fluctuate with large amplitude are
identified as ``Marginal''.

Evolutionary histories of a standard set of the hydrodynamic simulations 
are shown in Figure \ref{fig-HDevol},
where we plot the variations of maximum surface densities over 
time\footnote{Our realization of initial density perturbations
has the maximum density
$\Sigma_{\rm max}/\Sigma_0 = 1 + 3.5\epsilon_0$ at $t=0$.}.
The evolution of the RMS density fluctuation amplitude $\epsilon$
shows similar behavior.
We fix $q=1$, $n_J=5$, $\gamma=1.5$ and vary $\epsilon_0$ and $Q$.
It is clear that for all cases
the initial phase of evolution is governed by 
swing amplification, exhibiting rapid growth of the surface density.
While these initial swing amplifications in $Q\simgt0.9$ models become
saturated at $t/\torb\sim0.6-0.9$ from 
shear kinematics, nonlinear collapse of shearing wavelets 
immediately and continuously follows
the initial swing amplifications when $Q\simlt0.8$ (models H06 and H07).
Fourier analyses of the density distributions at each time when
the shearing box becomes exactly periodic reveal that 
the dominant mode during the swing amplification phase of these models has 
$\ky=1(2\pi/L)$. From the linear theory, Figure \ref{fig-KGam} shows
the total amplification of modes for $Q=1.1$ and 1.5 and $q=1$ as
a function of $K_y$, where $\ky=K_yn_J(2\pi/L)$. The maximum
amplification predicted is for $K_y=0.2-0.3$ corresponding to
$\ky=(1-1.5) \times (2\pi/L)$ for $n_J=5$, in good
agreement with the simulations\footnote{Recall that $\ky(2\pi/L)^{-1}$
can only take on integer values in the simulations.}.
As noted above, the growth times and the amplification magnitudes 
from the numerical simulations are listed in the seventh
and eighth columns of Table \ref{tbl-1}. 
For these columns we also list in parentheses the termination times $t_c$ 
and corresponding growth magnitudes for swing amplification of 
the $K_y=0.2$ modes in the linear theory, determined by integration of
equations (\ref{lcon})-(\ref{lmag}) from $t=0$ to $t=t_c$ 
(i.e.\ approximately half of the full swing interval).
For $1\leq Q \leq 1.4$,
if a mode were allowed to grow from $t=-t_c$, its 
amplification factor would be $\sim3-10$ times greater 
than that of a half swing. The contribution from the initial modes 
with $\kx <0$ that undergo a full swing amplification in the simulation
is very small, because the steep power spectrum sets the 
amplitudes of these $\kx\neq0, \ky\neq0$ modes as small compared to
the amplitudes of the corresponding $\kx=0$, $\ky=0$ modes.

While the fluctuation amplitudes vary slowly and continuously with the
parameter $Q$ during the initial swing phase,
the subsequent nonlinear evolution is critically dependent on $Q$,
exhibiting ``threshold'' behavior.
For the time being, we concentrate our discussion on 
the $\epsilon_0=10^{-2}$ cases.
When $Q \simlt 0.8$, as in model H06 and H07, the density enhancement
from swing is so large that shearing wavelets produce overdense
filaments, which become gravitationally 
unstable along their length, even before swing amplification ends. 
The filaments collapse and
fragment further into pieces at $t/\torb\sim0.5$. 
Model H06, for instance, forms six fragments with each mass in the range of
$\sim2-7\%$ of the total mass, which can be compared with
a characteristic Jeans mass of 4\% for $n_J=5$. 
The fragments subsequently collect background material and collide with 
each other to form larger clumps each of mass $\sim8-28\%$ 
at $t/\torb\sim0.8$.
When $Q$ is intermediate, as in models H08 ($Q=0.9$) and H09 ($Q=1.0$),
on the other hand, the growth of surface density from swing amplification
is not enough to initiate the collapse of the shearing wavelets,
and thus parallel fragmentation does not occur. 
Since the swing-amplified perturbations modify the initial
background velocity field significantly, the filamentary structures can also
move in the radial direction, and may collide with other
filaments that are moving in the opposite direction. 
The mutual gravitational forces between 
the moving filaments expedite physical collisions.  
With their larger surface density, the filaments in model H08 collide
more violently than those in model H09, quickly leading to the formation of 
four bound clumps at $t/\torb\sim1.0$. Model H09 experiences
five successive collisions of wavelets to form one big clump
at $t/\torb\sim2.2$, as shown in Figure \ref{fig-HDimage}.

When $Q\simgt1.1$, swing amplification is so mild
that it produces neither immediate parallel fragmentation of filaments,
nor physical collision of filaments followed by fragmentation. 
In this case, the transient swing 
amplification saturates into an ``intermediate'' 
state with many independent perturbations
with small radial wavelengths. When $Q\simgt1.3$, 
the amplitude of perturbations at the end of the swing phase is quite low, 
and thus the perturbed surface density remains always in the linear
regime throughout the evolution, showing no further growth beyond the 
intermediate-saturation state.
If $1.3>Q\simgt1.1$, however, swing amplification 
casts the system into a pivotal state where shearing
wavelets nonlinearly interact with each other (but not involving 
physical collisions). 
While the linear phase is characterized by the steady, 
kinematic increase of the radial wavenumber (verified directly by
examination of Fourier amplitudes),
nonlinear interactions among different modes in this pivotal state
can change the radial wavenumber dynamically, 
causing new small-$|\kx|$ modes to be produced.
These newly excited, small-$|\kx|$ modes 
are subject to swing amplification, and grow further.
If the nonlinear feedback is strong enough,  
the result of this ``rejuvenated swing'' is the formation of 
gravitationally bound condensations (see below).
With only moderate feedback, however, 
the models H10 and H11 (with $Q=1.1$ and 1.2) have
only fluctuating density fields with order-unity amplitudes,
with strong shear preventing clumps
from forming up to the limit of our integration.

Obviously, the $\epsilon_0=10^{-3}$ models are more stable than
their $\epsilon_0=10^{-2}$ counterparts, mostly because the shear instability
is only transient. When $\epsilon_0=10^{-3}$, $Q\simgt1.1$ models
remain stable, never entering a nonlinear stage. Model H01 ($Q=0.9$) becomes
unstable via the direct collisions of filaments, while nonlinear
feedback allowing rejuvenated swing amplification makes model 
H02 ($Q=1.0$) eventually unstable.
In spite of the different paths to the formation of gravitationally bound
objects,
Figure \ref{fig-HDevol} suggests that reduction of $\epsilon_0$ 
by a factor of 10 does not significantly alter 
the critical $Q$-value that discriminates between collapsing and 
non-collapsing simulation outcomes;
the nonlinear stability of self-gravitating, shearing disks 
is relatively insensitive to the initial perturbation amplitude.
The value $Q_c$, above which gravitationally bound objects
do not form, is in the range $\sim 1.1-1.3$ for these unmagnetized models.

\citet{elm91} studied cloud formation via the combination of
Parker, thermal, and gravitational instabilities, and argued that
gravitational instability of higher density filaments first grown from 
the swing mechanism could form discrete clouds.
He termed this parallel fragmentation process
the ``secondary'' gravitational instability.
Our simulations indicate that there are two additional processes 
that could potentially lead to gravitationally bound condensations:
physical collisions of filaments, and rejuvenated swing 
amplification following nonlinear feedback creating small-$|\kx|$ modes. 
In what follows, we shall use {\it secondary instability}
to include all three routes mentioned above.
The ninth column $\tau_{\rm 2nd}$ 
in Table \ref{tbl-1} lists the time when filaments start to fragment or 
collide, or secondary swing amplifications occur. 
We will continue the discussion of secondary instabilities
in \S6.3.

Figure \ref{fig-HDimage} shows 
comparative density snapshots at three different epochs, 
for the $\epsilon_0=10^{-2}$ models H09 with $Q=1.0$  (left column)
and H13 with $Q=1.4$ (right column).
Density structure is shown in logarithmic scale.
It is found that at the end of initial swing amplification
the potential vorticity $\xi\equiv |\nabla\times\mathbf{v}
+2\mathbf{\Omega}|/\Sigma$ is
maintained with its initial equilibrium value $\xi_0=(2-q)\Omega/\Sigma_0$
within $\sim4\%$ except the regions where shocks form,
confirming that in the absence of viscosity and magnetic fields,
potential vorticity is conserved (e.g., \cite{hun64,gam96,gam01}).
(Here, $\Omega$ is the angular velocity of the grid center at $x=y=0$.)
Small deviations of $\xi$ from $\xi_0$ are due to our non-isentropic 
initial density perturbations.
A consequence of this potential vorticity conservation in
steady flow is the modification
of the local epicycle frequency as $\kappa=\kappa_0(\Sigma/\Sigma_0)^{1/2}$
\citep{bal85, dwa96}, allowing for the correlation
between local shear rate $q_1\equiv -\Omega^{-1}dv_y/dx$ and surface density 
as $q_1=2-(2-q)\Sigma/\Sigma_0$, when the motion is predominantly azimuthal.
Nonsteady and/or radial flows would alter this correlation, and
both models H09 and H13 (with $q=1$) exhibit 
$q_1 \sim 1.6 -0.6 \Sigma/\Sigma_0$. That is, the local azimuthal shear
rate decreases with increasing surface density similarly to 
but not exactly following the prediction
for steady, purely-azimuthal flow.
Since the streamlines emanating from the shock fronts begin to
be mixed up as shearing wavelets
start to nonlinearly interact, the potential vorticity is not
conserved over the entire domain at the end of the simulations. 
It is apparent from Figure \ref{fig-HDimage} that when $Q=1.4$, 
perturbations do not grow enough to become nonlinearly unstable,
but instead become essentially sheared sonic oscillations.
At $t/\torb=3$, the oscillation amplitude of density
is $\sim 12\%$ of the background density and the radial
wavenumber has been increased by shear up to $\kx\simeq 19(2\pi/L)$. 
When $Q=1.0$, on the other hand, 
successive collisions of nonlinear 
sheared patches induced by gravity 
eventually form a condensation whose mass 
is 28\% of the total mass in the simulation domain.

We explore the effect of $\gamma$ on the simulation outcome
first by comparing model H09 ($\gamma=1.5$) with models H14 and H18 
(with $\gamma=1$ and 2, respectively), as
displayed  in Figure \ref{fig-HDaux}a for the $Q=1$ models. 
Although the initial evolution of surface density is independent of $\gamma$, 
higher $\gamma$ can reduce the effect of self-gravity
once density grows sufficiently, lowering
the saturation level of the initial swing amplification.
For example, the saturated density at the end of the swing
phase in model H18 with $\gamma=2$ is
15\% smaller than that of model H09 with $\gamma=1.5$.
Subsequent nonlinear interaction 
among wavelets is less active than in the $\gamma=1.5$ case as well. 
Nonlinear feedback is strong enough 
to increase the density in the $\gamma=2$ model up
to 7 $\Sigma_0$ at $t/\torb=1.5$, but a pressure bounce resulting
from the (unphysically) 
stiff equation of state combines with the background shear velocity
field to prevent formation of a condensation.
On the other hand,  in model H14 with $\gamma=1$ -- just as for the 
$\gamma=1.5$ model H09 -- 
shearing wavelets start to collapse before the swing amplification ends, and 
thermal pressure gradients are insufficient to halt gravitational runaway. 
As the collapse proceeds, fragmentation occurs 
(our resolution here [256$^2$] is not adequate to follow late-time 
evolution of fragments).
Models H11 to H13 with $\gamma=1.5$ (shown in Fig.\ \ref{fig-HDevol}b)
similarly may be compared with their $\gamma=1$ counterparts,
H15 to H17 (shown in Fig.\ \ref{fig-HDaux}a).
In Model H15 with $Q=1.2$ and $\gamma=1$, the saturation level of
surface density is higher than model H11 with $\gamma=1.5$, 
but not enough to be unstable immediately. 
Instability is deferred until $t/\torb\sim 2$,
when much more vigorous nonlinear feedback drives the system into
gravitational collapse. As $Q$ increases further, no secondary
instability occurs; 
even with $\gamma=1$, a local disk with $Q \geq 1.3$ is
stable to forming self-gravitating condensations.  Overall, we conclude 
that the $Q$ threshold for nonlinear gravitational instability is relatively
insensitive to the value of $\gamma$, decreasing by at most $\approx 0.2$ 
as $\gamma$ increases from 1 to 1.5.

To assess the extent to which our local approximation may affect
the dynamics, we have performed a number of simulations in which
we vary the size of the local box.
In \S3, we argued that the local
model should be acceptable for the study of linear MSA because the maximum
amplification is achieved at intermediate wavenumbers. 
We test whether the nonlinear outcome of MSA is indeed insensitive to the box 
size by repeating models H09 and H11 (with $n_J=5$) as models
H20 and H21 in domains four times as large in area (with $n_J=10$).
These $n_J=10$ models can harbor modes with wavelengths twice
as large as the maximum permitted in the $n_J=5$ models
under the periodic box boundary conditions.
If the inclusion of $\lambda=2L$ modes were to produce 
significant changes in simulation outcomes, our approach of 
investigating gravitational stability of galactic disks using a local model
would come into question. 
Reassuringly, results presented in Figure \ref{fig-HDaux}b show that 
the evolution of maximum surface density in $n_J=10$ models is 
in fact quite
similar to that in $n_J=5$ models\footnote{A simulation
box with $n_J$ can accommodate modes that
satisfy $K_y = j/n_J$, where $j$ is any positive integer between 1
and half the number of resolution elements in the 
$\mathbf{\hat{y}}$-direction. From Figure \ref{fig-KGam}a, one can
expect that $n_J=10$ models have an additional, small contribution 
from $K_y=0.1$ modes that are absent in 
$n_J=5$ models in the initial swing amplification.},
although the presence of $\lambda=2L$ modes in model H20
induce more violent late time evolution in collapsing objects, starting at
$t/\torb\sim 1.5$. As a result of its higher total mass, model H20 
forms more strongly bound clumps than model H09, 
but in terms of instability criteria, $n_J=10$ models
yield the same result as $n_J=5$ models.
These (and similar) experiments confirm that the local model is 
an acceptable tool to study stability of disks, provided the box size 
is large enough to contain the dominant growing modes.

Finally, we have confirmed that our standard resolution of
256$^2$ is adequate for the current purpose by comparing models
at varying resolution. Figure \ref{fig-HDaux}c shows, for example,
the comparison of models H11 with identical models at twice and half
the numerical resolution (H22, H23). 
Since $\kx$ is expected to increase linearly with time,
we need to make sure that nonlinear secondary instability 
is not an artificial result arising from insufficient late-time resolution
of high-$\kx$ modes.
The results shown in Figure \ref{fig-HDaux}c
indicate that the secondary instability is authentic, occurring 
at the same time ($t/\torb\sim 2.1$) regardless of the resolution. 
In fact, during the time interval $0.7<t/\torb<2.1$,
most of power remains below $\kx = 10(2\pi/L)$ for these models;  we find 
similarly that for other models in which secondary instability develops,
$\kx \simlt 10(2\pi/L)$ modes during the saturated intermediate state.
This suggests that nonlinear (or perhaps quasilinear) interactions of 
wavelets are continuously feeding power back into the low-$|k_x|$ regime,
preventing the secular kinematic increase of radial wavenumbers that would
otherwise occur absent such interactions (see Fig. \ref{fig-test}).

\subsection{Magnetohydrodynamic case}

To study the effect of the magnetic field on the nonlinear 
evolution of local shearing
disks, we have performed a number of shearing box simulations in which we 
include a purely toroidal initial field with 
strength given by $\beta=10$ or 1 (see eq.\ [\ref{beta}] for normalization).
The parameters and evolutionary characteristics of the simulations, 
each designated by the prefix M and 
a model number, are listed
in Table \ref{tbl-2}. 
For all of these MHD simulations, a 256$^2$ grid was used.
When shear is relatively strong ($q\sim 1$), as described in \S3, 
the presence of a magnetic field tends to play a stabilizing role 
in the linear phase of amplification. 
In the previous section, we showed that nonlinear secondary instabilities
can destabilize a hydrodynamic system that would
otherwise be ultimately stable.
The importance of secondary instabilities in a magnetized system may 
differ from the hydrodynamic case, and thus 
the primary question we address in this section is 
how the eventual fate of a system is modified by 
the inclusion of a magnetic field.

We plot the evolution of maximum surface density 
in Figure \ref{fig-MHD10evol} for $\beta=10$ cases (models M01-M14)
and in Figure \ref{fig-MHD1evol} for $\beta=1$ cases (models M15-M26), 
respectively. As expected from the linear theory, 
the magnetic pressure from embedded mean toroidal fields  
reduces the effect of self-gravity along the radial direction, 
causing MSA to cease earlier, and thus reducing corresponding 
initial amplification factors by, e.g., 
37\% for $\beta=10$ and 90\% for $\beta=1$, compared to the 
the unmagnetized case with $Q=1.0$.
Magnetic tension resists the Coriolis force, reducing
the amplitude of the perturbed radial velocities, but this effect is
not significant if shear is strong {\it and} the initial field
strength is relatively weak ($\beta \geq 1$).
Unlike in the axisymmetric case,
the simple replacement of $Q$ with $Q_M\equiv Q(1+1/\beta)^{1/2}$ does not
accurately represent the effects of magnetic fields in MSA
compared to swing amplification.

As in the hydrodynamic cases, 
the models with $\epsilon_0=10^{-2}$ and $Q\simlt0.8$ (M05, M06 with 
$\beta=10$; M19, M20 with $\beta=1$) 
all experience parallel fragmentation of filaments before MSA terminates,
while gravitationally-induced collisions of filaments dominate
the nonlinear stage in slightly higher-$Q$ models such as 
M07 and M08 (respectively with $Q=0.9$, 1.0 and $\beta=10$),
and M21 (with $Q=0.9$ and $\beta=1$). 
The comparison of the evolutionary histories of $Q=1$ models M08 
(in Fig.\ \ref{fig-MHD10evol}) and H09 (in Fig.\ \ref{fig-HDevol})
shows that embedded magnetic fields make 
evolution in the aftermath of collisions more violent, 
by removing the constraint of the potential vorticity conservation.
As a result, model M08 rapidly forms four bound clumps 
within $t/\torb\sim 1.6$, 
whereas model H09 forms a clump at $t/\torb\sim 2.2$, as explained before.
Because MSA yields a lower amplification of perturbations for 
smaller-$\beta$ models, however, model M22 (with $Q=1.0$ and $\beta=1$) 
does not show physical collisions of shearing wavelets 
to form condensations. Instead, model M22 becomes nonlinearly unstable
through a secondary swing amplification.

In intermediate-$Q$ models in which parallel fragmentation
and collisions of filaments do not occur,
shearing wavelets interact nonlinearly after the initial MSA phase ends, 
as in the intermediate-$Q$ hydrodynamic models.
With magnetic fields included, however,
these nonlinear interactions are much stronger,
more easily leading to formation of condensations via ``rejuvenated'' MSA. 
The best examples for nonlinear magnetic destabilization are
found in models M10 (with $Q=1.2$ and $\beta=10$) and 
M23 (with $Q=1.1$ and $\beta=1$). The corresponding unmagnetized models
(H11 and H10, respectively) are only marginally stable.
Snapshots of density and field configurations
of these unstable magnetized M10 and M23 models, at three selected times, are 
respectively shown in Figures \ref{fig-MHD10image} and
\ref{fig-MHD1image}, where we also display for comparison
stable magnetized models M13 and M25 with slightly larger $Q$.
Although the saturation level of surface density fluctuations from
the initial MSA is lower than
from unmagnetized swing amplification with same $Q$,
magnetic field lines that connect shearing wavelets enhance
nonlinear interaction and
limit the growth of $\kx$ below $8(2\pi/L)$ for models M10 and M23.
Small-$|\kx|$ modes are replenished by nonlinear 
interactions, and then grow further through secondary MSAs to 
dominate the evolution. 
The left-middle snapshots in Figures \ref{fig-MHD10image} 
and  \ref{fig-MHD1image} clearly show the
density structures associated with secondary swing amplification;
notice that the pitch angles in these snapshots are larger (indicating 
smaller $k_x$) than
the pitch angles in the upper (earlier) frames.
The condensations in model M10 formed as a consequence of 
secondary instability have 13\%, 8\%, and 5\% of the total mass
from left to right,
while model M23 has two condensations with mass each 24\% and 10\%
from left to right. Notice, from model M10 in Figure \ref{fig-MHD10image},
the characteristic signature of prograde rotation in the collapsed
condensations evident from the wrapping of embedded field lines. 

In \S6.1, we showed from hydrodynamic simulations that unmagnetized,
isentropic flows indeed conserve the potential vorticity $\xi$;
i.e., they maintain a strict linear relationship between vorticity and
surface density along a given streamline.
The presence of magnetic fields must, however, destroy potential
vorticity conservation (e.g., \cite{gam96}). 
With relatively weak magnetic fields ($\beta=10$), model M10
at the end of swing amplifications still exhibits a weak correlation
between vorticity and surface density, giving
$\xi=(1\pm 0.2)\xi_0$. In model M23 with $\beta=1$, however, 
vorticity does not correlate with surface density at all
throughout the evolution; the local shear rate at any point is
totally independent of the surface density.
Unlike the situation within spiral arms for pure hydrodynamics, therefore,
the local shear rate within spiral arms for an MHD flow need not be
a decreasing function of the surface density, depending on field alignment.

Experiments with different $\gamma$, with histories presented 
in Figure \ref{fig-MHDaux}a, demonstrate that
$\gamma$ only affects the strongly nonlinear phase, 
and does not generically alter the ultimate simulation outcome.
In particular, for $\gamma=1$ and $1.5$, there is no measured
difference (i.e. $<0.1$) 
in the gravitational runaway threshold levels $Q_c$.
Figure \ref{fig-MHDaux}b compares evolutionary histories with varying
simulation box sizes. In general the initial growth of surface density 
is slightly larger in $n_J=10$ models than in $n_J=5$ cases, and differences
of the growth factors between the models with $\beta=1$ are slightly
larger than in hydrodynamic cases.
This is because, as indicated by Figure \ref{fig-KGam}, the relative
importance of $K_y=0.1$ modes (that fit in $n_J=10$ but not $n_J=5$ boxes) 
to $K_y=0.2$ modes (that fit in both boxes) is larger when $\beta=1$
than when $\beta=\infty$. Nevertheless,
the simulation box size makes negligible difference to 
the computed nonlinear stability criterion, again supporting 
the appropriateness of a local
model for the study of disk stability. 
Finally, we have performed a set of simulations (models M35 to M38)
taking the 2D Burgers power spectrum 
$\langle |\Sigma_k|^2\rangle \propto k^{-3}$
as an initial density perturbation. The results are shown in 
Figure \ref{fig-MHDaux}c together with models M22 to M25, which
start with a 2D Kolmogorov spectrum. 
Although the 2D Burgers spectrum
leads to $\sim4-13\%$ higher saturation levels, 
the overall evolutionary behaviors, especially
final outcomes of models M35 to M38, are essentially indistinguishable from
those of model M22 to M25. 
This confirms that the gravitational instability criteria obtained 
from numerical simulations are independent of the adopted initial
perturbation spectra, as long as the power decreases as 
the wavenumber increases.

Even with embedded magnetized fields, models having relatively large $Q$ 
are unable to reach the regime where
nonlinear feedback is strong enough to induce a secondary
instability. With $\epsilon_0=10^{-2}$, we find critical $Q$ values 
$Q_c=1.4$ for $\beta=10$ and $Q_c=1.2$ for $\beta=1$.
These thresholds are subject to slight change if initial perturbation
amplitudes vary; Figures \ref{fig-MHD10evol} and
\ref{fig-MHD1evol} show that the $Q_c$ values increase by $\simlt 0.3$ 
as $\epsilon_0$ increases by a factor ten.  This behavior
attests to the fact that intermediate
saturation is caused not by nonlinear effects
but by the transient nature of MSA. 
Compared to the hydrodynamic critical value $Q_c=1.3$,
the higher critical value $Q_c=1.4$ in the $\beta=10$ sub-thermally 
magnetized system indicates that
stronger nonlinear feedback renders it more unstable,
while the lower critical value $Q_c=1.2$ in $\beta=1$ models,
indicative of greater stability, arises from
a lower growth in the initial MSA phase.

\subsection{Routes to structure formation}

In the previous two subsections, we showed from the numerical
simulations that whether magnetic fields are present or not, 
disks with $Q$ smaller than a critical value can form 
gravitationally bound structures, while perturbations in 
relatively large-$Q$ disks do not grow enough to produce
a gravitational runaway.
We also showed that gravitational runaways, when present,
occur as a consequence of secondary instabilities working
on the filaments or shearing wavelets that have already been
amplified by MSAs. The secondary instabilities include
three different processes:
(1) immediate parallel fragmentation of filaments, 
(2) gravitationally-induced collisions of sheared patches, and 
(3) rejuvenated swing amplification.
In this section we use specific examples to distinguish and
illustrate each of the secondary instabilities.

Figure \ref{fig-route} compares the development of structure via the
three different secondary instabilities. 
The left column in Figure \ref{fig-route} 
illustrates the parallel fragmentation process for model 
M05 (with $Q=0.7$ and $\beta=10$). The first frame of left column 
exhibits three filaments resulting from MSAs. 
The rightmost filament fragments into three discrete
clumps (marked by square, triangle, and circle) 
at about $t/\torb\sim 0.53$, while the leftmost filament
with smaller surface density produces only one clump (marked by diamond)
at $t/\torb\sim0.59$, 
as a result of the Jeans instability operating along the filaments.
A relatively low-density clump condensed from the central filaments
undergoes a collision with the remnant of the leftmost filament,
becoming a very high-density entity after $t/\torb \sim 0.65$.
\citet{elm91} semi-analytically calculated growth factors for this 
parallel fragmentation instability 
and showed that they are very sensitive to $Q$ and 
have a sharp threshold near $Q\sim1$, which is in good agreement with
the results of our simulations. 

Collisionally-induced structure formation is exemplified in the center column 
of Figure \ref{fig-route}, where we display four snapshots for model
M07 (with $Q=0.9$ and $\beta=10$). A larger $Q$-value than in model M05
inhibits the parallel-fragmentation process discussed above, 
and thus model M07 still retains 
filamentary structure at $t/\torb=0.51$.  Comparison of
the first two frames in center column shows that $\kx$ is almost unchanged
during this time interval, indicating that the velocity fields are
significantly modified from the initial values by both MSAs 
and gravitational interactions between the filaments;
as a consequence, $\kx$ does not secularly increase in time. 
Note that at $t/\torb=0.51$,
$\Sigma_{\rm max}\simeq3\Sigma_0$ and $v_{x,\rm max}\simeq0.2v_{0,\rm max}$, 
thus the radial motions of filaments are significant.
The two filaments surrounded by boxes in the first frame are moving
in radially opposite directions, and experience a physical collision with
each other at $t/\torb=0.65$. 
The collision, accelerated by the mutual gravity between filaments, produces
a new filament of higher density ($\Sigma\sim 5\Sigma_0$ at $t/\torb=0.80$).
The central part of the newly born filament is in turn 
Jeans unstable, developing into a bound condensation, as marked by
circle in the last frame of center column.
Later, sheared patches further collide with each other to produce three more
clumps ($t/\torb>1.19$).
As we explained in \S6.2, the presence of {\it weak} magnetic fields permits
stronger post-collisional collapse by breaking potential vorticity 
conservation; {\it strong} magnetic fields
could however inhibit physical collisions from taking place at all.

Finally, the right column of Figure \ref{fig-route} 
(see also Fig.\ \ref{fig-MHD1image}) demonstrates how 
rejuvenated swing amplification develops for model M23 (with $Q=1.1$
and $\beta=1$).
Since $Q$ is relatively large, initial MSAs in model M10 do not 
amplify perturbations to the
level where fragmentation or physical collisions of sheared filaments 
occur. Still, enhanced density is in the nonlinear regime, with
$\Sigma_{\rm max} \sim 1.3\Sigma_0$ when MSAs saturate.
Individual wavelets tend to keep shearing out, but their nonlinear 
interactions restrain the shift to larger $\kx$ of the overall power 
spectrum.  At $t/\torb=1.59$, the maximum power
of model M23 is achieved at $\kx=7(2\pi/L)$, and $\kx\simgt10(2\pi/L)$
modes have almost negligible power.
Continued nonlinear interactions among shearing wavelets can feed fresh
small-$|\kx|$ modes ($t/\torb\sim 1.91$)
that then undergo additional swing amplifications.
Wavefronts of the principal ``rejuvenated'' swinging wavelets are 
indicated by dotted lines; 
these are not readily perceived until they grow significantly 
($t/\torb > 2.32$),
because the power of these small-$|\kx|$ modes is initially small.
Operating on an already nonlinear background, 
the rejuvenated swing amplification efficiently produces 
shearing patches of high surface density, especially when
magnetic fields are present.
The maximum surface density at $t/\torb=2.48$
is $\Sigma_{\rm max}=2.4\Sigma_0$. Overdense regions collect 
ambient material and suffer collisions with each other, finally to 
appear as bound condensations (see Fig.\ \ref{fig-MHD10image}).

Although the above three types of the secondary instabilities are 
independent processes, in some cases they 
cooperate with each other to drive model disks into an ultimate
gravitational runaway. For example, in models H07, M08, and M20, 
some fragmenting filaments undergo mutual collisions, while
other filaments experience only parallel fragmentation.
In models M09 and M21, on the other hand, gravitationally-induced
collisions of sheared patches are not initially strong enough to produce
immediate fragmentation, but high-power small-$|\kx|$ modes
generated by the collisions grow very rapidly as
they swing around. Most generally, bound cloud formation may 
result from a combination of 
parallel fragmentation, physical collisions, and rejuvenated swing
amplification, all of which work on sheared spiral wavelets previously
amplified by MSAs.

\subsection{Higher amplitude perturbations and other effects}

In the preceding sections, we have presented simulations of the
growth of structure from instabilities of low-amplitude 
($\epsilon_0=10^{-3}$ or $10^{-2}$) perturbations. Real
galaxies, however, may have significantly larger
perturbations.

One way to address this is by initiating simulations similar to those
in \S6.1-6.2, but with larger initial amplitudes. When we
perform such simulations with $\epsilon_0=0.1$ and $\gamma=1.5$, 
we find that the perturbations, aided by
nonlinear effects acting from the outset of simulations,
are so easily amplified that hydrodynamic disks with
$Q\simlt1.2$ experience gravitational runaways within $\sim0.4\,\torb$.
The maximum surface density in larger-$Q$ models also grows to
a few times the mean density,
but the background shear and the strong pressure gradients 
associated with both
large $Q$ values and the stiff equation of state ($\gamma=1.5$) 
cause overdense regions in shearing wavelets to bounce back. 
Bouncing wavelets experience physical collisions with others
moving in the opposite direction, but with high $Q$,
bound condensations do not form.
When we use an isothermal equation of state, on the other hand,
there is essentially no pressure bounce for models with
$Q\simlt 1.9$, and overdense regions collapse in runaway fashion.
Still, $Q\simgt2$ models remain stable throughout the evolution.
With magnetic fields included, we observe the similar evolutionary
behavior to hydrodynamic cases, although 
magnetized collisions following pressure bounce form 
bound condensations more easily, and 
the critical values for the runaway collapse with $\gamma=1$ 
are slightly smaller, giving $Q_c=1.8-1.9$.

Does this imply that the critical $Q$ values in real galaxies should be
closer to 2 than the
estimates in \S6.1-6.2? We think not, because the initial 
conditions of the adopted power spectrum shape are unrealistic for 
actual large-amplitude perturbations. Perhaps more realistic
``initial'' conditions would be those occurring in the 
saturated intermediate state that develops from our 
low-amplitude simulations. These
conditions are characterized by perturbations of amplitude 
$\sim10-20\,$\% of the background density,
concentrated on radial scales $\sim 0.1\,L$ (our model has $n_J=5$, so from 
eq.\  [\ref{Ln}] with typical parameters these scale are less than a kpc). 
For these intermediate states, we 
can measure the effective $Q$ values for those models that lie
at the boundary of susceptibility to secondary instabilities, 
taking allowance for the random perturbed velocity field, and
modified local shear and local sound speed produced by the growth of 
perturbations.
We find that the models (H11, M12, and M23) 
which show gravitational runaway have
$Q_{\rm eff} \leq 1.4$. The typical time required for the nonlinear
interactions to produce the low-$|\kx|$ modes needed to seed the
rejuvenated swing appears to be in the range of $\sim(1.5-2)\,\torb$.
Thus, we conclude  that the critical values of $Q$ to produce
gravitational runaway
from ``natural'' large-amplitude perturbations (other than spiral arms) 
are comparable to the
critical values estimated in \S\S6.1, 6.2 to grow from
small-amplitude perturbations.

A related issue concerns the detailed specification of the 
perturbations.
For the simulations presented in this paper, we impose perturbations
only on surface density, 
and initially kept the other variables uniform and constant.
When evolved with an adiabatic equation of state,
these isothermal perturbations give self-gravity a slight imbalance over
the pressure gradient force, causing overdense regions to attract
more material from the beginning of simulations.
When we instead perturb both surface density and thermal energy in 
accordance with an isentropic initial condition, 
we find that the system undergoes an initial adjustment 
while trying to find the most unstable modes for MSAs. 
This relaxation stage involves a reduction of the
density fluctuation amplitudes, sometimes affecting the destiny of
the model's dynamical evolution.
For example, a model that has the same parameters as model M23 except
initially isentropic perturbations experiences 
an initial decrease of density fluctuation amplitude by a factor of $\sim1.6$, 
and thus remains stable until the end of the simulation. 
Larger initial perturbation amplitudes lead to a larger ``relaxation''
reduction in fluctuating density amplitudes.
This suggests that if isentropic rather than isothermal perturbations
occur, then the critical $Q$ values become slightly smaller.
Equivalently, it implies that a slightly larger $\epsilon_0$ is needed to  
induce secondary instabilities if perturbations are isentropic 
rather than isothermal. Because only models with already-marginal $Q$
are subject to these differences, we do not consider the exact
specification of initial conditions (except for their scale)
crucial to the outcome.

\section{Discussion}

\subsection{Summary of model results}

In this paper, we have investigated both linear and nonlinear evolution of
self-gravitational instabilities arising from
nonaxisymmetric perturbations in local models of differentially rotating, 
magnetized, gaseous galactic disks. 
Our primary goals were to understand
the parametric dependences of these instabilities, and to determine
the conditions that ultimately lead to gravitational runaway.
The disk models are infinitesimally thin, and we adopt
an ideal gas equation of state with either an adiabatic or an isothermal
pressure-density relation.
We assume a uniform initial azimuthal magnetic field with 
$\beta=\cs^2/\vA^2$ (proportional to the gas-to-magnetic pressure ratio) 
characterizing the midplane field strength (see eq.\ [\ref{beta}]); 
we treat the magnetic field scale height as a constant in space and time.
The local model we use incorporates the shear profile, tidal gravity, 
and Coriolis force arising from orbits in a smoothly-varying 
stellar + dark matter galactic gravitational
potential, but does not include features such as stellar spiral arms.
The relative importance of galactic rotational shear, self-gravity,  
and thermal pressure are described by
the Toomre $Q$ stability parameter (see eq.\ [\ref{TQ}]) and the 
Jeans number $n_J$ of the spatial scale under consideration
(see eq.\ [\ref{nJ}]). 
The background flow's shear rate is measured by $q\equiv -d\ln\Omega/d\ln R$.

Our linear-theory analysis (\S3) is aimed at exploring and delimiting 
the physical mechanisms for {\it initiating} self-gravitating
condensations in two-dimensional disks.
In \S3.3, we employ the shearing-sheet formalism
and integrate the resulting linearized equations over time. 
As long as $q\ne 0$, growth of any linear perturbation eventually saturates
owing to shear, so that a disturbance with a given (fixed) azimuthal 
wavelength can be characterized by the magnitude $\Gamma$ of its total
amplification over all time.  We extensively analyze the dependence of 
the linear-theory saturated-state amplification on $q$, $Q$, and $\beta$.

Our linear analysis shows that there exist two distinct
kinds of nonaxisymmetric instabilities (\S3.3). 
Regions with weak shear ($q\ll 1$) and/or strong magnetic fields ($\beta\ll 1$)
are susceptible to magneto-Jeans instabilities (MJI). For these modes
(generalizations of those analyzed by \citet{lyn66}),
the presence of magnetic fields is essential for instability:
tension forces transfer angular
momentum, thereby greatly reducing the stabilizing effect of 
epicyclic motions.
The stronger the magnetic fields, the more unstable low-shear regions 
become to nonaxisymmetric motions.
On the other hand, when shear is relatively strong ($q\sim 1$) and
the mean magnetic field moderate or weak ($\beta \simgt 1$), 
the growth of perturbations arises 
from magnetically-modified swing amplification (MSA), with 
a stabilizing role played by magnetic pressure.
MSA is a generalization for MHD flows of the process originally
studied by GLB.
When $q\ll 1$, as needed for the MJI when galactic values of 
$\beta \simgt 1$ prevail, the slow increase of the local radial 
wavenumber gives disturbances plenty of time
to attain huge amplification. When $q\sim 1$, as needed for the MSA, 
growth of perturbations is more moderate. 
The inclusion of magnetic fields lowers the amplification factor
in the linear swing mechanism.

We also investigate the MJI using the coherent
wavelet approach (\S3.2), which allows us to obtain 
a closed-form dispersion relation (see eq.\ [\ref{sol4}]),
an instantaneous instability criterion (see eq.\ [\ref{kJ}]), 
and approximate expressions for amplification magnitudes 
(see eqs.\ [26]).
The amplification magnitudes obtained with this simplified approach
agree well with the exact integrations of the linearized 
equations.

To determine the eventual fate of growing condensations -- in particular,
whether or not bound/collapsing clumps eventually form -- we turn to
numerical simulations.
In our nonlinear evolution studies, we fix $n_J=5$ for our
standard simulation box (see eq.\ [\ref{Ln}]) and allow $q$, $Q$,
and $\beta$ to vary so as to represent different disk models.
Initially, we introduce small-amplitude density perturbations
by adding either white noise (for MJI) or Gaussian random 
noise with a 2D Kolmogorov power spectrum (for MSA)
and follow the time evolution of each disk model up to four orbital times. 

In \S5, we present the results of simulations with $q=0.1$, corresponding
to nonlinear evolution of MJI modes. With such low shear, 
perturbations grow almost exponentially to collapse 
within $\sim 2$ orbital times. 
The growth in the linear stage is predominantly along the mean
field direction, and its rate is independent of the adiabatic index $\gamma$. 
When $\gamma\simgt 1.6$, thermal pressure halts the collapse
and produces bound clumps.

In \S 6, we present the results of simulations with $q=1$, i.e. flat 
rotation curves. 
These represent our study of the eventual outcomes from MSA.
Because of the kinematic effects of shear, MSAs can grow for
only a limited time before 
saturating at $t/\torb\sim 0.7$ ($\sim 10^8$ yrs) 
(unless strong nonlinear effects enter earlier).
Disks with sufficiently small $Q<Q_c$ eventually collapse or form 
gravitationally bound clumps via
nonlinear secondary instabilities (\S6.3) 
operating on sheared spiral wavelets that have already been amplified
by MSAs. These secondary instabilities include
parallel fragmentation of filaments, gravitationally-induced 
collisions of nonlinear
sheared patches, and ``rejuvenated'' swing amplification.
 
Which type of the secondary instability dominates depends on $Q$. 
In disks with $Q\simlt0.8$, 
parallel fragmentation instabilities dominate,
while intermediate-$Q$ disks ($Q\thickapprox0.9-1.0$ depending on $\beta$)
preferentially exhibit collisionally-induced structure formation.
When $Q$ is slightly larger ($1.1\simlt Q <Q_c$ for $\beta\simgt10$ or
$1.0\simlt Q <Q_c$ for $\beta=1$), the intermediate-state density 
enhancement from MSAs is too low to produce parallel fragmentation or
physical collisions, but a second phase of swing amplification occurs. 
In ``rejuvenated'' swing,
nonlinear interactions among sheared wavelets can supply fresh 
small-$|\kx|$ modes that then undergo sufficient swing amplification to 
produce gravitational runaway. 

Our simulations show that embedded magnetic fields can strengthen nonlinear 
feedback, and accelerate post-collision collapse.
The time required for structure formation tends to be shorter for lower $Q$,
but for $Q>1$, ultimate instability occurs not before
1.6 orbital times ($4\times 10^8$ yrs). 
The masses of collapsing regions or pressure-supported clumps are 
typically $\sim 10^7\,\Msun$. 

Models with sufficiently large $Q>Q_c$ have low saturated-state 
density fluctuations and thus
are not susceptible to secondary instability.
We find that the critical values of $Q$ that discriminate between
the final outcomes, with gravitational runaway absent for $Q\geq Q_c$,
are $Q_c=1.3$ for $\beta=\infty$, $Q_c=1.4$ for $\beta=10$, and 
$Q_c=1.2$ for $\beta=1$,
when the root mean square density fluctuations are initially $\epsilon_0=1\%$.
These critical $Q$ values are found to be
insensitive to the simulation box size and the computational resolution,
and slightly dependent (at a 10-20\% level) on $\epsilon_0$ and the 
adiabatic index $\gamma$.

As noted above, the computed values of $Q_c$ from our nonlinear,
nonaxisymmetric simulations are relatively insensitive to the magnetic
field strength.  In linear (nonaxisymmetric) theory, the corresponding
result is the relatively weak dependence of the amplification
magnitude for MSA on $\beta$ (see Figs.
\ref{fig-qGam}-\ref{fig-KGam}).  At the most basic level, this can be
understood from the physics of linear swing amplification: most of the
growth occurs during the ``open spiral'' phase of the swing ($\kx(t)$
near zero), when density ridges form near-radial spokes and the
perturbed flow velocities are primarily azimuthal.  Since the
background magnetic fields are themselves azimuthal, flow in the open
phase is nearly along field lines and hence relatively unaffected by
magnetic forces.  For comparison, {\it linear, axisymmetric} modes
have instability thresholds at $Q_c=(1+ \beta^{-1})^{-1/2}$, or 1,
0.95, and 0.71 for $\beta=\infty,$ 10, and 1, respectively.  Our
finding that the nonlinear, nonaxisymmetric $Q_c$ is smallest for $\beta=1$ 
may reflect the reduction in amplification by radial magnetic pressure
gradients during the more strongly leading/trailing phases of the swing.
Our finding that the nonlinear, nonaxisymmetric $Q_c$ is larger for 
$\beta=10$ than for $\beta=\infty$ is indicative of the role magnetic fields
play in ``rejuvenating'' swing by enhancing feedback to low-$\kx$ modes.  
We note that this result could not have been predicted solely 
from the linear analysis:  Figure \ref{fig-Qbeta} shows that 
the amplification magnitude $\Gamma$ increases with increasing 
$\beta$, for linear MSA.

\subsection{Application to inner galaxies}

One of the most interesting results of our study is seeing how
magnetic fields dramatically alter dynamics
in a weak-shear environment.
The underlying physical reason for this is that magnetic tension
reduces the stabilizing effect from Coriolis forces, and
thus in the limiting case of very strong fields
makes shear instability essentially the same as the 2D Jeans 
instability in the azimuthal direction (cf. \cite{elm87a}).
This MJI could, at least partly,  explain the active star formation 
observed towards the central parts of galaxies where rotation curves are 
nearly solid body and the star formation rate is generally orders of
magnitude higher than in outer disks (e.g., \cite{ken98}).  Nuclear 
starburst activity -- or even bulge formation in a 
dark matter halo without a central cusp -- may represent 
the extreme of this phenomenon.

Other physical processes that might influence the nuclear 
star-formation activity 
include molecular gas content, external disturbances from nearby or 
interacting galaxies, and nuclear gas transport via inner Lindblad 
resonance and bars (cf. \cite{jog00}). 
However, there is intriguing evidence that starburst activity is closely
linked to the slope of the rotation curve. For example, the starburst 
galaxy M82 has vigorous star formation in the rigidly rotating part, 
and weak star formation beyond the turnover point of the rotation 
curve \citep{tel91}. 
\citet{ken93} also found that the starburst region in NGC 3504 corresponds to
the linearly rising portion of the rotation curve. 
\citet{ken93} estimated $Q$ near 0.9 for the center of this
galaxy, which does not rule out the possibility of axisymmetric 
gravitational instability provided $\beta>4$. The corresponding growth time
for axisymmetric modes
is $\zeta^{-1}\sim\Omega^{-1}$ from equation (\ref{sol4})
with $\beta=\infty$.
On the other hand, if there are significant embedded magnetic
fields ($\beta<4$) in the azimuthal direction that preclude
{\it axisymmetric} instabilities, then {\it nonaxisymmetric}
MJIs can account for the growth of gas complexes. 
The coherent wavelet solution (\ref{sol4}) yields
the growth time for nonaxisymmetric modes of 
$\tau_{\rm grow} \equiv \zeta_{\rm max}^{-1} =
(0.5-0.7)\,\Omega^{-1}$ for $Q=0.9$, $q=0$, and $\beta=1-10$
(with higher $\beta$ giving larger  $\tau_{\rm grow}$).
For $\Omega \sim (500-1000) \,{\rm km\,s^{-1}\,kpc^{-1}}$ 
observed in the central few hundred parsecs of NGC 3504,
this amounts to $\sim(5-14)\times 10^5$ yrs.

MJIs could also potentially explain the vigorous star formation activity 
near the inner
regions of the late-type spiral galaxies M33 and NGC 2403,
where rotation curves are rising slowly with radius.
\citet{ken89} and \citet{mar01} found that 
the inner parts of both galaxies have gas surface 
densities well below the minimum required to meet the 
$Q$-threshold criterion empirically determined from outer-galaxy star 
formation.  \citet{mar01} suggest, however, that if one applies an alternative
criterion for instability in weak-shear regions proposed by 
\citet{elm93b} and \citet{hun98} for irregular galaxies, 
the star formation in the interiors of M33 and NGC 2403 can be explained.  
Here, we relate to and generalize this ``shear criterion'' based on our 
results for MJI.

Similarly to \citet{elm93b}, we posit that instability significant enough
to produce active star formation corresponds to sufficient total growth 
of the MJI before shear causes saturation.  With
$\Gamma_{\rm max}$ the log of the total amplification of linear perturbations, 
$\Gamma_{\rm max}> \Gamma_c$ for some $\Gamma_c$ thus defines an instability
threshold.  If magnetic fields are very strong, 
($(c_s/v_A)^2\equiv \beta \ll 1$), growth at the Jeans rate 
$\approx \pi G \Sigma/c_s$ persists for a time $(q \Omega)^{-1}$.  Equation
(26a) would thus imply instability when 
\begin{equation}\label{QA1}
Q_A\equiv \frac{q \Gamma_c Q}{2\sqrt{1-q/2}}=
\frac{2 \Gamma_c A \cs}{\pi G \Sigma_0}
< 1,{\rm \;\;\;\;for\;\beta\ll1},
\end{equation}
where $A\equiv q \Omega /2$ is Oort's shear parameter. The shear criterion
of \citet{elm93b} corresponds to equation (\ref{QA1}) with $\Gamma_c=1.25$.

When magnetic fields are moderate or weak ($c_s/v_A \simgt 1$), the MJI growth
rate is reduced below the Jeans rate, and our equation (26b) would then 
imply instability for weak-shear regions 
where $q\simlt 0.7\beta^{-1/3}Q^{-1}$ provided
\begin{equation}\label{W_cond}
Q_w \equiv \left(\frac{q \Gamma_c \cs}{1.6 \vA}\right)^{1/2}Q < 1.
\end{equation}
Our {\it nonlinear} simulations of the MSA indicate that the threshold for
gravitational runaway occurs at $Q$ such that $\Gamma$ for {\it linear} 
wavelets would be near unity (see Tables 1,2).  If the critical 
amplification $\Gamma_c$ for MJI is similar, then equation (\ref{W_cond}) 
would predict instability provided $(q c_s/v_A)^{1/2} Q \simlt 1.3$.  The
threshold surface density from equation (\ref{W_cond}) is generally larger
than from equation (\ref{QA1}), but smaller than would nominally be required
for MSA (which however {\it cannot} occur in low-shear regions).  
In particular, if we use the observational parameters for NGC 2403 from 
\citet{mar01}, 
instability under the condition (\ref{W_cond}) with $\Gamma_c\simgt 1$ is 
predicted only in the very innermost regions; to have
instability over the whole inner region would require $\Gamma_c<1$.  
Since $\Gamma<1$ would correspond to 
relatively weak growth of perturbations, it seems likely that instead 
either the steep part of
the rotation curve extends to larger radius than presently estimated 
(current measures suffer from low resolution) -- allowing  
$\Gamma_c\simgt 1$, or that effects from spiral arms
are important for explaining the high star formation rate in much of the 
inner portion of this galaxy.

\subsection{Application to outer galaxies}

Since the bulk of disk material in spiral galaxies orbits following
a flat rotation profile (e.g., \cite{sof99}), $q\sim 1$ is 
a good representation for local shear outside of
the central regions.
Magnetic fields in spiral galaxies have a mean field component
of typically a few $\mu$G \citep{bec96}, 
corresponding to $\beta\sim1$ to 10. Therefore, MSA modes
are probably more important than MJI modes in initiating the growth
of self-gravitating structures in extended disks of spiral galaxies -- 
{\it outside} of lowered-shear spiral arms.

The two main characteristics of linear MSA modes are the limit to initial
growth of imposed perturbations from the shearing of wave crests, 
and the stabilizing effects of embedded magnetic fields.
Our numerical simulations show that if $Q$ is not less than unity
and initial multi-kpc scale perturbations are $\sim1$\%, 
the density enhancement at 
$t\sim 10^8$ yrs, when MSAs are fully developed and kinematically 
saturated, is not large enough to produce gravitationally
bound structures.
The presence of magnetic fields reduces the growth of density 
perturbations by MSA
even further, so that when $\beta=1$, only $Q\simlt 0.8$ models 
can achieve sufficient density to collapse right away.
Although subsequent nonlinear interactions among shearing wavelets
could give an extra boost for $1\simlt Q\simlt 1.4$, 
leading to the formation of bound 
condensations, the whole process
takes more than $\sim4\times 10^8-10^9$ yrs, longer than   
the available timescale of $10^8$ yrs for the growth
of perturbations without being disturbed by random supernova events,
formation of OB associations, and larger-scale density waves
\citep{elm87a, elm93a}.
This suggests that MSAs in disks 
can produce observed GMAs or HI superclouds in spiral galaxies 
where shear is relatively high and $Q$ is large,
only with the aid of additional agents.  We discuss some potential 
additional effects below.

In this study, we idealized the multiphase gaseous medium using 
inviscid monolayers, so that our model
disks do not incorporate the destabilizing effect of viscosity
potentially associated with mutual collisions of cloudlets having 
large mean free paths (cf. \cite{gam96}).
By making a thin-disk approximation, our model
disks overestimate the self-gravity at the disk midplane
(cf. \cite{too64}); a correction for finite thickness would 
decrease the critical $Q$-values. 
A more significant drawback associated
with the simplified geometry is that 2D simulation models cannot
capture the potential consequences of magnetorotational
instabilities (MRIs; e.g., \cite{bal98}) and 
the Parker instability \citep{par66}
on cloud formation. MRIs, 
which exist only in 3D systems, are
known to generate MHD turbulence in accretion disks and may also
contribute to large-scale turbulence in galactic disks 
\citep{sel99}. Because gravitational instabilities in 
outer disks are primarily limited by shear, and MRIs require weak enough 
magnetic fields that shear is not significantly suppressed, MRIs may primarily
affect growth timescales rather than $Q$ thresholds.
Parker instability has often been invoked in the formation
of molecular clouds along spiral arms where shear is reduced 
(e.g., \cite{bli80}; see
discussion in \cite{elm95a} and references therein).  Outside of spiral 
arms, the primary role of Parker instability might be in helping to 
seed large-scale
perturbations which then become self-gravitating via MSA.
To understand the full dynamical impact of these instabilities combined
with self-gravity,
it is desirable to expand the current work into 3D, also 
making allowance for a multiphase ISM.

In the present study, we considered structure formation only in a gaseous
component and followed its linear and nonlinear evolution under its
own gravity. Real galaxies, of course, consist of both gaseous and stellar
disks. Although the velocity dispersion of stars is larger than the 
gas sound speed typically by an order of magnitude, the contribution
of the stellar part to the dynamical evolution of the combined system 
for axisymmetric linear modes is almost
comparable to that of the gaseous part if the gas 
fraction is $\sim10-25\%$ \citep{jog84a}. 
\citet{elm95b} and \citet{jog96} extended the work of \citet{jog84a}
and showed that 
the gravitational coupling of these two components makes
the combined two-fluid system 
more unstable than each disk alone. 
\citet{jog92} studied nonaxisymmetric swing amplification in a
two-fluid system and found that unlike the situation for
axisymmetric instability
(in which gas and stars respond coherently with the same growth rates),
the lower velocity dispersion of the gas in the nonaxisymmetric 
instability allows for a larger amplification in gas than in stars.
More extensive study of the effect of a ``live'' stellar component
on the linear and nonlinear outcome of MSA in a 
gaseous disk represents an important direction for future research.
Qualitatively, it is clear that 
the inclusion of stellar gravity would tend to increase
the critical $Q$ values and shorten the time for the instability to
form gravitationally bound objects.

While the linear theory yields 
MSA amplification factors that vary continuously with respect to $Q$,
our numerical simulations show that a small change in $Q$ near the
critical value results in dramatically different final
states, owing to nonlinear interactions of shearing wavelets.
The evidence for critical $Q$ values in deciding the outcome of the 
nonaxisymmetric instabilities is suggestive of
the observationally-inferred
threshold for determining global star formation in spiral galaxies.
Although bound structures may take some time to form, the critical
values ($Q_c\sim1.2-1.4$ depending on $\beta$) obtained from our
simulations are similar to the
observational threshold value $Q_c\sim1.4$ very recently reported by
\citet{mar01}.  This quantitative agreement is certainly encouraging, 
but of greater importance is simply the concrete demonstration that disks 
which are linearly stable ($Q>1$) nevertheless show robust threshold behavior
for self-gravitational runaway under a variety of conditions -- as anticipated
by empirical models.  

The empirical threshold model
predicts radial trends in star formation by employing
azimuthally-averaged gas surface densities and rotation curves,
while in fact azimuthal variations of gas surface densities and the shear rate
due to gas streaming motions  are substantial.
Given the strong spatial association of giant cloud complexes and active
star formation with spiral density waves, the azimuthally-averaged 
star formation threshold 
model is evidently oversimplified.
A straightforward extension of the ``axisymmetric'' star
formation threshold model would be to take account of spiral
structure, with active star formation at some radius 
expected provided that the
local value of $Q$ within arms at that radius is smaller than some
threshold. This extension tends to bring observed thresholds closer
to our estimates and also reduces growth timescales, as we explain next.

Long-lived global spiral density waves \citep{lin64,lin66} may have a 
dramatic effect
on small-scale structure formation, because the enhanced density 
in wave crests makes the local $Q$ value and 
shear rate smaller than in the interarm regions.
In the hydrodynamic theory, 
$Q_s=Q_0(\Sigma_s/\Sigma_0)^{-0.5}$ for an isothermal gas,
and the local shear gradient is $q_s=2-(2-q)\Sigma_s/\Sigma_0$,
where the ``$s$'' and ``0'' subscripts denote respectively arm and interarm
values \citep{bal85, elm94}. These scalings 
suggest that it would not be difficult to obtain $Q_s\simlt 0.7$ and 
$q_s \sim 0$
for typical arm-interarm contrasts of a few to several in spiral galaxies.
Naive application of our present analyses to spiral arm crests with
such small $Q$ and $q$ local conditions suggests that 
collapse of arm gas could proceed very rapidly via MJI modes,  
forming condensations within one Jeans time scale (a few
times $10^7$ yrs).
The observed strong association of giant cloud complexes with 
spiral arms (e.g., \cite{kenney97}) supports the idea of their triggered 
formation mechanism by spiral density waves.
GMAs even in flocculent galaxies appear to be associated with
weak, near-infrared arms (e.g., \cite{tho97a, tho97b}).

Although the above extrapolations of (unmagnetized) linear theory and 
indications from
observations are highly suggestive, the detailed nonlinear MHD evolution
within arms has yet to be explored theoretically. Linear analysis
 suggests that additional stabilizing and destabilizing
terms may be important when the background has large density variations;
i.e., a strong arm \citep{bal88}.
It is unknown, however, how the epicycle frequency (or, equivalently
local shear rate) varies inside arms when
the conservation of the potential vorticity is broken due to
magnetic fields.
Do phase correlations of the background density and velocity gradients
in the arm still determine the character of instabilities, as suggested by
\citet{bal88} when $B=0$?
Numerical computations with explicit inclusion of spiral arm
potentials are required
to address these and related questions, extending the effort to 
understand the highly complex problem of cloud and star formation in spiral 
galaxies.

\acknowledgements

We gratefully acknowledge J.\ M.\ Stone for providing the original ZEUS 2D 
code and many helpful insights, and are thankful to C.\ F.\ Gammie for 
sharing his shearing-sheet Poisson solver and expert comments on the 
manuscript.  We also thank S. Vogel and B. Elmegreen for constructive 
comments, and are grateful to an anonymous referee for a 
detailed and insightful report.  This work was supported by NASA grants 
NAG 53840 and NAG 59167.

\input{table1}

\input{table2}

\clearpage
\begin{figure}
\epsscale{1.}
\caption{Total amplification magnitude $\Gamma$ of linear gravitational
instabilities in a thin shearing disk with an embedded azimuthal
magnetic field. $Q=1.3$ and $K_y\equiv \ky/k_J=0.5$ are adopted. Solid lines
are drawn from the direct numerical integration of the linearized
shearing sheet equations, while dotted lines are the results
of the coherent wavelet solutions; when $q\ll 1 $ or $
\beta\ll 1$, corresponding to the condition for MJI, 
these two results are in good agreement with each other.
For $0.6<q<2$ and $\beta \simgt 1$, MSA prevails, with solutions 
following the unmagnetized swing amplification result (heavy curve).
\label{fig-qGam}}
\end{figure}

\begin{figure}
\epsscale{1.}
\caption{Parametric dependence on $Q$ and $\beta$ of total 
linear-theory amplification magnitude $\Gamma$ for the 
case of $K_y=0.2$ and $q=1$.
The numbers labeling heavy contours correspond to $\Gamma$.
The interval of the light contours is given by $\Delta\Gamma =0.1$.
The discontinuity in $\Gamma$ at $\beta_c\sim0.1-1$ is indicative of the two
different mechanisms responsible for instabilities: MJI for $\beta<\beta_c$ and
MSA for $\beta > \beta_c$. See text for details.
\label{fig-Qbeta}}
\end{figure}

\begin{figure}
\epsscale{1.}
\caption{Total linear-theory amplification magnitude $\Gamma$ as a function 
of the dimensionless azimuthal wavenumber $K_y\equiv k_y/k_J$.
For $\beta\simgt 1$, $\Gamma$ peaks at $K_y\sim0.15-0.4$ with
smaller- and larger-$K_y$ regions stabilized by epicyclic motion
and MHD waves, respectively. 
For $\beta\ll 1$, stabilization by the epicyclic motion occurs
at $K_y\simlt 0.1\beta$. Discontinuities in $\beta=1$ curves
show that MSA (MJI) applies for smaller (larger) $K_y$ modes.
\label{fig-KGam}}
\end{figure}

\begin{figure}
\epsscale{1}
\caption{A test run with a $128\times 128$ resolution for MSA 
of a single nonaxisymmetric mode. 
Solid lines obtained from the numerical
simulation are in good agreement with open circles taken from
the results of linear analysis. $Q=2$, $n_J=2.5$, $q=1$, $\gamma=1.5$,
and $\beta=1$ are adopted and
sinusoidal perturbations with an amplitude
($\Sigma_1$, $U_1$, $v_{x,1}$, $v_{y,1}$, $B_{x,1}$, $B_{y,1}$) =
$10^{-4}$(1, 108, -6.16, 1, 1, 6) are initially imposed. 
\label{fig-test}}
\end{figure}

\begin{figure}
\epsscale{1.}
\caption{Evolution of maximum surface density for nonlinear MJIs. 
$Q=1.5$, $n_J=5$, and $\beta=1$ are adopted, with initial white noise 
perturbations of amplitude $10^{-3}\Sigma_0$. Since $q=0.1$ gives almost an 
exponential growth of perturbations, its evolution is not much
different from the $q=0$ case, causing overdense regions to collapse
within $\sim 2$ $\torb$. When $\gamma\simgt 1.6$, 
thermal pressure can support the collapsed regions and form bound objects.
\label{fig-MJI-devol}}
\end{figure}

\begin{figure}
\epsscale{1.16}
\caption{An example of MJIs with $Q=1.5$, $n_J=5$, $q=0.1$, $\beta=1$, and
$\gamma=2$. Snapshots at $t=2$ ({\it left}) and 4 
({\it right}) orbits show
surface density in logarithmic grey scale, perturbed velocity vectors, and 
magnetic field lines. The arrow above each frame measures the amplitudes
of velocity vectors, and the numbers in the grey scale bars correspond
to $\log\;\Sigma/\Sigma_0 $.
\label{fig-MJI-image}}
\end{figure}

\begin{figure}
\epsscale{1.}
\caption{Time evolution of maximum surface density in pure 
hydrodynamic simulations for models H01 to H13.
\label{fig-HDevol}}
\end{figure}

\begin{figure}
\epsscale{1.2}
\caption{Comparative snapshots of density structure in logarithmic 
grey scale for model H09 with $Q=1.0$ (left) and for model
H13 with $Q=1.4$ (right).
The numbers in the grey scale bars correspond to $\log \,\Sigma/\Sigma_0 $.
\label{fig-HDimage}}
\end{figure}

\begin{figure}
\epsscale{1.}
\caption{Effects on the evolution of hydrodynamic simulations of 
(a) adiabatic index $\gamma$, (b)
simulation box size (equal to $\lambda_Jn_J$), and (c) 
numerical resolution. 
All models have $\epsilon_0=10^{-2}$ initially.
\label{fig-HDaux}}
\end{figure}

\begin{figure}
\epsscale{1.}
\caption{Time evolution of maximum surface density in MHD simulations
with $\beta=10$ (models M01 to M14).
\label{fig-MHD10evol}}
\end{figure}

\begin{figure}
\epsscale{1.}
\caption{Time evolution of maximum surface density in MHD simulations
with $\beta=1$ (models M15 to M26).
\label{fig-MHD1evol}}
\end{figure}

\begin{figure}
\epsscale{1.2}
\caption{Comparative evolution of magnetized disks with different $Q$
but with the same magnetization of $\beta=10$ .
Left (for model M10 with $Q=1.2$) and right (for model
M13 with $Q=1.5$) columns show density structures in logarithmic 
grey scale with overlaid magnetic field lines.
The numbers in the grey scale bars correspond to $\log\, \Sigma/\Sigma_0 $.
\label{fig-MHD10image}} 
\end{figure}

\begin{figure}
\epsscale{1.2}
\caption{Comparative evolution of magnetized disks with different $Q$
but with the same magnetization of $\beta=1$ .
Left (for model M23 with $Q=1.1$) and right (for model
M25 with $Q=1.3$) columns show density structures in logarithmic 
grey scale with overlaid magnetic field lines.
The numbers in the grey scale bars correspond to $\log\, \Sigma/\Sigma_0 $.
\label{fig-MHD1image}}
\end{figure}

\begin{figure}
\epsscale{1.}
\caption{Effects on the evolution of MHD simulations with $\beta=1$ 
of (a) adiabatic index $\gamma$, (b)
simulation box size (given by $L=\lambda_Jn_J$), and (c) the initial
power spectrum of perturbations.
\label{fig-MHDaux}}
\end{figure}

\begin{figure}
\epsscale{1.05}
\vspace{-0.8cm}
\caption{Examples of three different kinds of the secondary instabilities.
The numbers in the grey scale bars correspond to $\log\, \Sigma/\Sigma_0 $.
Left: Filaments formed by MSAs experience parallel fragmentation 
to form gravitationally bound structures in model M05. 
Various symbols identify and follow each clump.
Center: Collisions of nonlinear wavelets produce a high-density filament
that collapses in Model M07.
Boxes in the first frame highlights the portions of the wavelets that
are about to undergo a collision. Circled in the last frame is
the region where a bound clump forms.
Right: Rejuvenated swing amplification produces a filament that
subsequently collapses, in Model M23. The secondary-MSA wavefronts are
represented by dotted lines.
For details, see text.
\label{fig-route}}
\end{figure}

\end{document}

%% file: table1.tex
\begin{deluxetable}{cccccccccc}
\tabletypesize{\footnotesize}
\tablecaption{Parameters of Hydrodynamic Simulations with Strong Shear. 
\label{tbl-1}}
\tablewidth{0pt}
\tablehead{
\colhead{ \begin{tabular}{c} Model \\ (1)  \end{tabular} }         &
\colhead{\begin{tabular}{c} $Q$    \\ (2) \end{tabular} }           &
\colhead{\begin{tabular}{c} $n_J$  \\ (3) \end{tabular} }         &
\colhead{\begin{tabular}{c} $\gamma$ \\ (4)\end{tabular} }      &
\colhead{\begin{tabular}{c} $\epsilon_0$ \\ (5) \end{tabular} }    &
\colhead{\begin{tabular}{c} Grid \\ (6)\end{tabular}  }          &
\colhead{\begin{tabular}{c} $\tau_{\rm 1st}$\tablenotemark{a,b,c} \\ (7)      
        \end{tabular} }   &
\colhead{\begin{tabular}{c} $\Gamma_{\rm 1st}$\tablenotemark{c,d} \\ (8) 
        \end{tabular} } &
\colhead{\begin{tabular}{c} $\tau_{\rm 2nd}$\tablenotemark{b} \\ (9) 
        \end{tabular} }  &
\colhead{\begin{tabular}{c} Result \\ (10)\end{tabular}  }        
}
\startdata
H01 & 0.9 & 5 & 1.5 &$10^{-3}$& 256$\times$256 & 0.88(0.87) & 2.36(2.30) & 0.8 & Unstable \\
H02 & 1.0 & 5 & 1.5 &$10^{-3}$& 256$\times$256 & 0.79(0.78) & 1.95(1.95) & 2.0 & Unstable \\
H03 & 1.1 & 5 & 1.5 &$10^{-3}$& 256$\times$256 & 0.72(0.71) & 1.63(1.63) & $>$4&   Stable \\
H04 & 1.2 & 5 & 1.5 &$10^{-3}$& 256$\times$256 & 0.70(0.65) & 1.38(1.38) & $>$4&   Stable \\
H05 & 1.3 & 5 & 1.5 &$10^{-3}$& 256$\times$256 & 0.65(0.60) & 1.18(1.19) & $>$4&   Stable \\
 & & & & & & & & & \\ 
H06 & 0.7 & 5 & 1.5 &$10^{-2}$& 256$\times$256 &......(1.11)&......(3.35)& 0.5 & Unstable \\
H07 & 0.8 & 5 & 1.5 &$10^{-2}$& 256$\times$256 &......(0.97)&......(2.75)& 0.6 & Unstable \\
H08 & 0.9 & 5 & 1.5 &$10^{-2}$& 256$\times$256 & 0.82(0.87) & 1.99(2.30) & 0.7 & Unstable \\
H09 & 1.0 & 5 & 1.5 &$10^{-2}$& 256$\times$256 & 0.78(0.78) & 1.85(1.95) & 1.1 & Unstable \\
H10 & 1.1 & 5 & 1.5 &$10^{-2}$& 256$\times$256 & 0.76(0.71) & 1.61(1.63) & 1.9 & Marginal \\
H11 & 1.2 & 5 & 1.5 &$10^{-2}$& 256$\times$256 & 0.69(0.65) & 1.37(1.38) & 2.1 & Marginal \\
H12 & 1.3 & 5 & 1.5 &$10^{-2}$& 256$\times$256 & 0.65(0.60) & 1.18(1.19) & $>$4&   Stable \\
H13 & 1.4 & 5 & 1.5 &$10^{-2}$& 256$\times$256 & 0.58(0.56) & 1.04(1.03) & $>$4&   Stable \\
 & & & & & & & & & \\ 
H14 & 1.0 & 5 & 1.0 &$10^{-2}$& 256$\times$256 &......(0.78)&......(1.95) & 0.6&    Unstable \\ 
H15 & 1.2 & 5 & 1.0 &$10^{-2}$& 256$\times$256 & 0.70(0.65) & 1.40(1.38) & 1.9 &    Unstable \\ 
H16 & 1.3 & 5 & 1.0 &$10^{-2}$& 256$\times$256 & 0.65(0.60) & 1.20(1.19) &$>$4 &   Stable \\
H17 & 1.4 & 5 & 1.0 &$10^{-2}$& 256$\times$256 & 0.58(0.56) & 1.05(0.03) &$>$4 &   Stable \\
 & & & & & & & & & \\ 
H18 & 1.0 & 5 & 2.0 &$10^{-2}$& 256$\times$256 & 0.76(0.78) & 1.78(1.95) & 1.1 &   Marginal \\
H19 & 1.2 & 5 & 2.0 &$10^{-2}$& 256$\times$256 & 0.69(0.65) & 1.36(1.38) & 2.1 &   Marginal \\
 & & & & & & & & & \\ 
H20 & 1.0 & 10 & 1.5 &$10^{-2}$& 256$\times$256 & 0.71(0.78)& 1.84(1.95) & 1.1 &  Unstable  \\
H21 & 1.2 & 10 & 1.5 &$10^{-2}$& 256$\times$256 & 0.69(0.65)& 1.40(1.38) & 2.1 &  Stable   \\
 & & & & & & & & & \\ 
H22 & 1.2 & 5 & 1.5 &$10^{-2}$& 512$\times$512 & 0.69(0.65) & 1.37(1.38) & 2.1 &   Marginal  \\
H23 & 1.2 & 5 & 1.5 &$10^{-2}$& 128$\times$128 & 0.69(0.65) & 1.36(1.38) & 2.1 &   Marginal \\ 
\enddata

\tablenotetext{a}{defined by the time at which $\epsilon\equiv
           <(\Sigma/\Sigma_0-1)^2>^{1/2}$ attains its first maximum}
\tablenotetext{b}{in the unit of the orbital period $\torb$}
\tablenotetext{c}{the values in the parentheses are from the linear analyses
            with $K_y=0.2$.}
\tablenotetext{d}{$\Gamma_{\rm 1st} = \log[\epsilon(\tau_{\rm 1st})
    / \epsilon_0] = \case{1}{2}\log[<(\Sigma-\Sigma_0)^2>_{\tau_{\rm 1st}}
    / <(\Sigma-\Sigma_0)^2>_0 ]$}

\end{deluxetable}

%% file: table2.tex
\begin{deluxetable}{cccccccccc}
\tabletypesize{\footnotesize}
\tablecaption{Parameters of Magnetohydrodynamic Simulations 
with Strong Shear. \label{tbl-2}}
\tablewidth{0pt}
\tablehead{
\colhead{\begin{tabular}{c} Model            \\ (1)\end{tabular}}&
\colhead{\begin{tabular}{c} $\beta$          \\ (2)\end{tabular}}&
\colhead{\begin{tabular}{c} $Q$              \\ (3)\end{tabular}}&
\colhead{\begin{tabular}{c} $n_J$            \\ (4)\end{tabular}}&
\colhead{\begin{tabular}{c} $\gamma$         \\ (5)\end{tabular}}&
\colhead{\begin{tabular}{c} $\epsilon_0$       \\ (6)\end{tabular}}&
\colhead{\begin{tabular}{c} $\tau_{\rm 1st}$\tablenotemark{a,b,c} \\ (7)
        \end{tabular} }   &
\colhead{\begin{tabular}{c} $\Gamma_{\rm 1st}$\tablenotemark{c,d} \\(8)
   \end{tabular} } &
\colhead{\begin{tabular}{c} $\tau_{\rm 2nd}$\tablenotemark{b}   \\(9)
   \end{tabular} } &
\colhead{\begin{tabular}{c} Result           \\(10)\end{tabular}}
}
\startdata
M01 & 10& 1.0 & 5 & 1.5 &$10^{-3}$& 0.75(0.78)&1.76(1.69)&2.4 & Unstable \\
M02 & 10& 1.1 & 5 & 1.5 &$10^{-3}$& 0.72(0.73)&1.48(1.42)&3.4 & Marginal \\
M03 & 10& 1.2 & 5 & 1.5 &$10^{-3}$& 0.65(0.65)&1.26(1.21)&$>$4&   Stable \\
M04 & 10& 1.3 & 5 & 1.5 &$10^{-3}$& 0.59(0.55)&1.10(1.05)&$>$4&   Stable \\
 & & & & & & & & &  \\ 
M05 & 10& 0.7 & 5 & 1.5 &$10^{-2}$&......(1.11)&......(2.96)&0.5 & Unstable \\
M06 & 10& 0.8 & 5 & 1.5 &$10^{-2}$&......(0.97)&......(2.44)&0.6 & Unstable \\
M07 & 10& 0.9 & 5 & 1.5 &$10^{-2}$& 0.71(0.86)&1.87(2.02)&0.9 & Unstable \\
M08 & 10& 1.0 & 5 & 1.5 &$10^{-2}$& 0.74(0.78)&1.70(1.69)&1.1 & Unstable \\
M09 & 10& 1.1 & 5 & 1.5 &$10^{-2}$& 0.72(0.73)&1.46(1.42)&1.7 & Unstable \\
M10 & 10& 1.2 & 5 & 1.5 &$10^{-2}$& 0.64(0.65)&1.26(1.21)&2.1 & Unstable \\
M11 & 10& 1.3 & 5 & 1.5 &$10^{-2}$& 0.59(0.55)&1.10(1.05)&2.5 & Unstable \\
M12 & 10& 1.4 & 5 & 1.5 &$10^{-2}$& 0.58(0.54)&0.97(0.91)&3.2 & Marginal \\
M13 & 10& 1.5 & 5 & 1.5 &$10^{-2}$& 0.57(0.52)&0.87(0.81)&$>$4&   Stable \\
M14 & 10& 1.6 & 5 & 1.5 &$10^{-2}$& 0.56(0.49)&0.78(0.72)&$>$4&   Stable \\
 & & & & & & & & &  \\ 
M15 & 1 & 0.8 & 5 & 1.5 &$10^{-3}$& 0.52(0.59)&1.46(1.42)&1.4 & Unstable \\
M16 & 1 & 0.9 & 5 & 1.5 &$10^{-3}$& 0.50(0.51)&1.26(1.23)&3.5 & Marginal \\
M17 & 1 & 1.0 & 5 & 1.5 &$10^{-3}$& 0.48(0.48)&1.09(1.07)&$>$4&   Stable \\
M18 & 1 & 1.1 & 5 & 1.5 &$10^{-3}$& 0.48(0.47)&0.95(0.95)&$>$4&   Stable \\
 & & & & & & & & &  \\ 
M19 & 1 & 0.7 & 5 & 1.5 &$10^{-2}$&......(0.69)&......(1.68)&0.6 & Unstable \\
M20 & 1 & 0.8 & 5 & 1.5 &$10^{-2}$&......(0.59)&......(1.42)&0.8 & Unstable \\
M21 & 1 & 0.9 & 5 & 1.5 &$10^{-2}$& 0.54(0.51)&1.27(1.23)&1.2 & Unstable \\
M22 & 1 & 1.0 & 5 & 1.5 &$10^{-2}$& 0.50(0.48)&1.10(1.07)&1.4 & Unstable \\
M23 & 1 & 1.1 & 5 & 1.5 &$10^{-2}$& 0.48(0.47)&0.97(0.95)&2.3 & Unstable \\
M24 & 1 & 1.2 & 5 & 1.5 &$10^{-2}$& 0.47(0.46)&0.87(0.85)&$>$4&   Stable \\
M25 & 1 & 1.3 & 5 & 1.5 &$10^{-2}$& 0.46(0.45)&0.78(0.76)&$>$4&   Stable \\
M26 & 1 & 1.4 & 5 & 1.5 &$10^{-2}$& 0.46(0.44)&0.71(0.69)&$>$4&   Stable \\
 & & & & & & & & &  \\ 
M27 & 1 & 1.0 & 5 & 1.0 &$10^{-2}$& 0.50(0.48)&1.10(1.07)&1.0 & Unstable \\
M28 & 1 & 1.1 & 5 & 1.0 &$10^{-2}$& 0.48(0.47)&0.97(0.95)&2.3 & Unstable \\
M29 & 1 & 1.2 & 5 & 1.0 &$10^{-2}$& 0.47(0.46)&0.87(0.85)&$>$4&   Stable \\
M30 & 1 & 1.0 & 5 & 2.0 &$10^{-2}$& 0.50(0.47)&1.09(1.07)&$>$4& Unstable \\
 & & & & & & & & &  \\ 
M31 & 1 & 1.0 & 10& 1.5 &$10^{-2}$& 0.51(0.48)&1.14(1.07)&1.3 & Unstable \\
M32 & 1 & 1.1 & 10& 1.5 &$10^{-2}$& 0.49(0.47)&0.99(0.95)&2.1 & Unstable \\
M33 & 1 & 1.2 & 10& 1.5 &$10^{-2}$& 0.48(0.46)&0.87(0.85)&3.7 &  Stable  \\
M34 & 1 & 1.3 & 10& 1.5 &$10^{-2}$& 0.47(0.45)&0.80(0.76)&$>$4&  Stable  \\ 
& & & & & & & & &  \\ 
M35\tablenotemark{e} & 1 & 1.0 & 5 & 1.5 &$10^{-2}$& 0.49(0.48)&1.14(1.07)&1.4 & Unstable \\
M36\tablenotemark{e} & 1 & 1.1 & 5 & 1.5 &$10^{-2}$& 0.48(0.47)&0.99(0.95)&2.4 & Unstable \\
M37\tablenotemark{e} & 1 & 1.2 & 5 & 1.5 &$10^{-2}$& 0.47(0.46)&0.89(0.85)&$>$4&   Stable \\
M38\tablenotemark{e} & 1 & 1.3 & 5 & 1.5 &$10^{-2}$& 0.46(0.45)&0.81(0.76)&$>$4&   Stable \\
\enddata
\tablenotetext{a}{defined by the time at which $\epsilon\equiv
           <(\Sigma/\Sigma_0-1)^2>^{1/2}$ attains its first maximum}
\tablenotetext{b}{in the unit of the orbital period $\torb$}
\tablenotetext{c}{the values in the parentheses are from the linear analyses
            with $K_y=0.2$.}
\tablenotetext{d}{$\Gamma_{\rm 1st} = \log[\epsilon(\tau_{\rm 1st})
    / \epsilon_0] = \case{1}{2}\log[<(\Sigma-\Sigma_0)^2>_{\tau_{\rm 1st}}
    / <(\Sigma-\Sigma_0)^2>_0 ]$}

\tablenotetext{e}{2D Burgers specturm is used to generate initial 
density perturbations}

\end{deluxetable}